\documentclass[11pt]{article}
\pdfoutput=1
\usepackage{booktabs}
\usepackage{jcapmod}
\usepackage{graphicx}
\usepackage{amsmath}
\usepackage{amsfonts}
\usepackage{amssymb}
\usepackage{color}
\usepackage{epsf}
\usepackage{array}
\usepackage{longtable}
\newcolumntype{L}[1]{>{\raggedright\let\newline\\\arraybackslash\hspace{0pt}}m{#1}}
\newcolumntype{C}[1]{>{\centering\let\newline\\\arraybackslash\hspace{0pt}}m{#1}}
\newcolumntype{R}[1]{>{\raggedleft\let\newline\\\arraybackslash\hspace{0pt}}m{#1}}
\usepackage{float}
\usepackage{pifont}
\usepackage{dcolumn}
\usepackage{bm}
\usepackage{epsfig}
\usepackage{amssymb,latexsym,mathrsfs}
\usepackage{hyperref}
\usepackage{etoolbox}
\newcolumntype{C}[1]{>{\centering\let\newline\\\arraybackslash\hspace{0pt}}m{#1}}

\def\beq{\begin{equation}}
\def\eeq{\end{equation}}
\def\be{\begin{equation}}
\def\ee{\end{equation}}
\def\bea{\begin{eqnarray}}
\def\eea{\end{eqnarray}}

\newcommand{\smallfoot}[1]{\let\thefootnote\relax\footnotetext{\scriptsize{#1}}}

\definecolor{darkgreen}{rgb}{0,0.5,0}
\definecolor{darkblue}{rgb}{0,0,0.75}
\definecolor{royalblue}{RGB}{0,0,128}
\definecolor{darkred}{RGB}{210,0,0}

\newcommand{\kms}{\mbox{km s$^{-1}$}}

\newcommand\degree       {{\ifmmode^\circ\else$^\circ$\fi}}  
\newcommand\arcm         {{\ifmmode {'\ }\else$'     $\fi} } 
\newcommand\arcs         {{\ifmmode{''\ }\else$''    $\fi} } 

\newcommand{\msun}{\mbox{$\rm M_{\odot}$}}

\newcommand{\lsun}{\mbox{$L_{\odot}$}~}

\def\aj{\rmfamily{AJ~}}           
\def\apj{\rmfamily{ApJ~}}         
\def\apjl{\rmfamily{ApJ~}}        
\def\apjs{\rmfamily{ApJS~}}       

\def\aap{\rmfamily{A\&A~}}        

\def\araa{\rmfamily{ARA\&A~}}     
\def\mnras{\rmfamily{MNRAS~}}     
\def\jcap{\rmfamily{JCAP~}}     
\def\prd{\rmfamily{Phys.~Rev.~D~}}
\def\prl{\rmfamily{Phys.~Rev.~Lett.~}}
\def\nat{\rmfamily{Nature~}}      
\def\ao{\rmfamily{Appl.~Opt.~}}   

\def\procspie{\rmfamily{Proc.~SPIE~}}

\begin{document}
\setcounter{page}{1} \baselineskip=14.5pt \thispagestyle{empty}

\bigskip\

\title{Science Impacts of the SPHEREx All-Sky Optical to Near-Infrared
  Spectral Survey II: \\
\vspace{1cm}
\emph{Report of a Community Workshop on the Scientific Synergies 
Between the SPHEREx Survey and Other Astronomy Observatories}
}

\abstract{SPHEREx, the Spectro-Photometer for the History of the Universe, Epoch of Reionization, and Ices Explorer, is a proposed NASA MIDEX mission selected for Phase A study pointing to a downselect in early CY2019, leading to launch in CY2023. SPHEREx would carry out the first all-sky spectral survey at wavelengths between 0.75 and 2.42 $\mu$m [with spectral resolution R=41], 2.42 and 3.82 $\mu$m [with R=35],  3.82 and 4.42 $\mu$m [with R=110], and 4.42 and 5.00 $\mu$m [with R=130]. At the end of its two-year mission, SPHEREx would obtain 0.75-to-5$\mu$m spectra of every 6.2$\times$6.2 arcsec pixel on the sky, with a 5-sigma sensitivity AB$>$19 per spectral/spatial resolution element.  SPHEREx would obtain spectra of every sources in the 2MASS PSC (1.2 $\mu$m, 1.6 $\mu$m, 2.2 $\mu$m) catalog to at least (40 $\sigma$, 60 $\sigma$, 150 $\sigma$) per spectral channel, and spectra with S/N $\geq$3 per frequency element of the faintest sources detected by WISE. More details concerning SPHEREx are available at \href{[Link]}{http://spherex.caltech.edu}. The SPHEREx team has proposed three specific science investigations to be carried out with this unique data set: cosmic inflation,  interstellar and circumstellar ices, and the extra-galactic background light.

Though these three scientific issues are undoubtedly compelling, they are far from exhausting the scientific output of SPHEREx.  Indeed, as Table~\ref{tab:statistics} shows, SPHEREx would create a unique all-sky spectral database including spectra of very large numbers of astronomical and solar system targets, including both extended and diffuse sources.   These spectra would enable a wide variety of scientific investigations, and the SPHEREx team is dedicated to making the SPHEREx data available to the scientific community to facilitate these investigations, which we refer to as Legacy Science.  To that end, we have sponsored two workshops for the general scientific community to identify the most interesting Legacy Science themes and to ensure that the SPHEREx data products are responsive to their needs.  In February of 2016, some 50 scientists from all scientific fields met in Pasadena to develop these themes and to understand their implications for the SPHEREx mission.  The results of this initial workshop are reported in Dor\'e et al., 2016.  Among other things, discussions at the 2016 workshop highlighted many synergies between SPHEREx Legacy Science and other contemporaneous astronomical missions, facilities, and databases.  Consequently, in January 2018 we convened a second workshop at the Center for Astrophysics in Cambridge to focus specifically on these synergies. This white paper, which contains substantial contributions from the participants, presents some of the highlights of the 2018 SPHEREx workshop.\footnote{\textcircled{c} 2018. All rights reserved.}   
}

\vspace{0.6cm}

\author{Olivier  Dor\'e$^{1,21}$, Michael W. Werner$^{1,21}$, Matthew L.\ N.\ Ashby$^{26}$, 
  Lindsey E. Bleem$^{36,37}$, Jamie Bock$^{21,1}$, Jennifer Burt$^{33}$,
  Peter Capak$^8$, Tzu-Ching Chang$^{1,21}$, 
  Jon\'as Chaves-Montero$^{36}$,
  Christine H. Chen$^{32}$,
  Francesca Civano$^{32}$,
  I. Ilsedore Cleeves$^{32}$,
  Asantha Cooray$^{28}$, Brendan Crill$^{1,21}$, 
  Ian J. M. Crossfield$^{39}$,
  Michael  Cushing$^4$, Sylvain de la Torre$^8$,
  Tiziana DiMatteo$^{25}$,
  Niv Dvory$^{38}$,
  Cora Dvorkin$^{42}$,
  Catherine Espaillat$^{22}$,
  Simone Ferraro$^{40}$,
  Douglas Finkbeiner$^{26}$,
  Jenny Greene$^{16}$,
  Jackie Hewitt$^{33}$,
  David W. Hogg$^{10}$,
  Kevin Huffenberger$^6$, 
  Hyun-Sung Jun$^{7}$,
  Olivier Ilbert$^9$,
  Woong-Seob Jeong$^7$,
  Jennifer Johnson$^{18}$,
  Minjin Kim$^{7}$,
  J.Davy Kirkpatrick$^8$,
  Theresa Kowalski$^1$,
  Phil Korngut$^{21}$,
  Jianshu Li$^{13}$,
  Carey M. Lisse$^{11}$,
  Meredith MacGregor$^{43}$,
  Eric E. Mamajek$^{1}$,
  Phil Mauskopf$^{30}$,
  Gary Melnick$^{26}$, 
  Brice M\'enard$^{14}$, 
  Mark Neyrinck$^{23}$,
  Karin \"Oberg$^{31}$,
  Alice Pisani$^{16}$,
  Jennifer Rocca$^{1}$,
  Mara Salvato$^{35}$,
  Emmanuel Schaan$^{40,41}$,
  Nick Z. Scoville$^{21}$,
  Yong-Seon Song$^7$,
  Daniel J. Stevens$^{18}$, 
  Ananth Tenneti$^{25}$,
  Harry Teplitz$^8$,
  Volker Tolls$^{26}$, 
  Stephen  Unwin$^{1}$,
  Meg Urry$^{34}$,
  Benjamin Wandelt$^{12,15,19,20}$, 
  Benjamin F. Williams$^{31}$,
  David Wilner$^{26}$,
  Rogier A. Windhorst$^{30}$, 
  Scott Wolk$^{26}$,
  Harold W. Yorke$^5$, 
  Michael Zemcov$^{27,1}$\\ 
\vspace{0.75cm}
\emph{Editors and corresponding authors: Olivier Dor\'e$^{1,21}$,
  Michael Werner$^{1,21}$ (\texttt{olivier.dore@caltech.edu, michael.w.werner@jpl.nasa.gov})}}

\affiliation{$^1$ Jet Propulsion Laboratory, California Institute of Technology, 4800 Oak Grove Drive, Pasadena, CA 91109, USA\\
$^2$ Universities Space Research Association, Stratospheric
Observatory for Infrared Astronomy, NASA Ames Research Center, MS 232-11, Moffett
Field, CA 94035, USA\\
$^3$ ASIAA, AS/NTU, 1 Roosevelt Rd Sec 4, Taipei, 10617, Taiwan\\
$^4$ Department of Physics and Astronomy, University of Toledo, 2801
W. Bancroft St., Toledo, OH 43606, USA\\
$^5$ USRA, SOFIA Science Center, Moffett Field, CA 94035-0001, USA\\
$^6$ Department of Physics, Florida State University, PO Box 3064350, Tallahassee, Florida 32306-4350, USA\\
$^7$ Korea Astronomy and Space Science Institute, Daejeon, 34055, Korea\\
$^8$ IPAC, Caltech, 770 S. Wilson Ave, Pasadena, CA 91125, USA\\
$^9$ Aix Marseille Univ, CNRS, LAM, Laboratoire d’Astrophysique de Marseille, Marseille, France\\
$^{10}$ Center for Cosmology and Particle Physics, Department of Physics, New York University, New York, NY 10003, USA\\
$^{11}$ JHU-APL, SES/SRE, Bldg 200/E206, 11100 Johns Hopkins Road,
Laurel, MD 20723, USA \\
$^{12}$ Center for Computational Astrophysics, Flatiron Institute, 162 5th Ave, New York City, NY 10010, USA\\
$^{13}$ Massachusetts Institute of Technology, Department of Physics, 70 Vassar Street, Bldg. 37-656, Cambridge, MA 02139, USA\\
$^{14}$ Dept. of Physics and Astronomy, Johns Hopkins University, 3400 N. Charles St., Baltimore, MD 21218, USA\\
$^{15}$ Institut d’Astrophysique de Paris (IAP), UMR 7095, CNRS UPMC, Universit\'e Paris 6, Sorbonne Universit\'e, 98bis boulevard Arago, F-75014 Paris, France\\
$^{16}$ Department of Astrophysical Sciences, Peyton Hall, Princeton University, Princeton, NJ 08544, USA\\
$^{17}$ Physics \& Astronomy Department, Vanderbilt University, Nashville, TN 37235, USA\\
$^{18}$ Department of Astronomy, The Ohio State University, 140 W. 18th Avenue, Columbus, OH 43210 USA\\
$^{19}$ Institut Lagrange de Paris (ILP), Sorbonne Universit\'e, 98bis boulevard Arago, F-75014 Paris, France\\
$^{20}$ Department of Physics and Astronomy, University of Illinois at Urbana-Champaign, 1002 W Green St, Urbana, IL 61801, USA\\
$^{21}$ California Institute of Technology, 1200 E. California Blvd, Pasadena, CA 91125, USA\\
$^{22}$ Department of Astronomy, Boston University, 725 Commonwealth Avenue 
Boston, MA 02215, USA\\
$^{23}$ Department of Theoretical Physics, University of the Basque Country UPV/EHU, 48080 Bilbao, Spain\\
IKERBASQUE, Basque Foundation for Science, 48013 Bilbao, Spain\\
$^{24}$ Dept. of Physics and Astronomy, Johns Hopkins University, 3400 N. Charles St., Baltimore, MD 21218, USA\\
$^{25}$ McWilliams Center, Carnegie Mellon University, Pittsburgh, PA 15213, USA\\
$^{26}$ Harvard Smithsonian CfA, 60 Garden St., Cambridge, MA 02138, USA\\
$^{27}$ Rochester Institute of Technology, Center for Detectors, 1 Lomb
Memorial Drive, Rochester, NY 14623, USA\\
$^{28}$ University of California Irvine, 4186 Frederick Reines Hall,
Irvine, CA 92697, USA\\
$^{30}$ Arizona State University, Department of Physics, P.O. Box
871504, Tempe, AZ 85287, USA\\
$^{31}$ Astronomy Department, Box 531580, University of Washington, Seattle, WA 98195, USA\\
$^{32}$ Space Telescope Science Institute, 3700 San Martin Dr., Baltimore, MD 21218, USA\\
$^{33}$ MIT Kavli Institute, Building 37, 70 Vassar Street, Cambridge, MA, USA\\
$^{34}$ Yale Center for Astronomy \& Astrophysics, Yale University, New Haven, CT 06520, USA\\
$^{35}$ Cluster of Excellence, Boltzmann Strasse 2, 85748 Garching, Germany\\ 
$^{36}$ Argonne National Laboratory, High-Energy Physics Division, 9700 S. Cass Avenue, Argonne, IL 60439, USA\\
$^{37}$ Kavli Institute for Cosmological Physics, University of Chicago, 5640 South Ellis Avenue, Chicago, IL 60637, USA\\
$^{38}$ McDonald Observatory, Department of Astronomy, The University of Texas at Austin, 
2515 Speedway, Austin, Texas 78712, USA\\
$^{39}$ Department of Physics, Massachusetts Institute of Technology, 77 Massachusetts Avenue, Cambridge, MA, USA\\
$^{40}$ Center for Cosmological Physics and Department of
Astronomy, University of California, Berkeley, CA 94720, USA\\
$^{41}$ Lawrence Berkeley National Laboratory, One Cyclotron Road, Berkeley, CA 94720, USA\\
$^{42}$ Harvard University, Department of Physics, Cambridge, MA 02138, USA\\
$^{43}$ Carnegie Department of Terrestrial Magnetism, 5241 Broad Branch Road NW, Washington, D.C. 20008
}

\maketitle

\section{The SPHEREx Mission Concept}
\label{sec:intro}

SPHEREx, the Spectro-Photometer for the History of the Universe, Epoch
of Reionization, and Ices Explorer is a proposed NASA MIDEX mission selected
for Phase A study before a downselect in late CY2018, leading to
launch in CY2023.  The Principal Investigator is Professor 
Jamie Bock of Caltech.  SPHEREx would carry out the first
all-sky spectral survey at wavelengths between 0.75 and 2.42 $\mu$m
with spectral resolution R=41, 2.42 and 3.82 $\mu$m with R=35,  3.82 and 4.42 $\mu$m with R=110, and 4.42 and 5.00 $\mu$m with R=130. At the end of its two-year mission, SPHEREx would obtain 0.75-to-5$\mu$m spectra of every 6.2$\times$6.2 arcsec$^{2}$ pixel on the sky, with a 5-$\sigma$ sensitivity AB$\simeq$19-19.5 per spectral/spatial resolution element (see Fig.~\ref{fig:lam_mag}). SPHEREx would measure the spectrum of every object in the 2MASS PSC (1.2 $\mu$m, 1.6 $\mu$m, 2.2 $\mu$m) catalog to at least (40 $\sigma$, 60 $\sigma$, 150 $\sigma$) per spectral channel.  Most objects in the WISE catalog would also be detected by SPHEREx spectroscopically, the faintest detected at $\simeq 3\sigma$ in each spectral channel (Fig.~\ref{fig:lam_mag}).

The SPHEREx science team has optimized the mission to address three scientific questions consistent with the three major themes of NASA’s astrophysics program.  In particular, SPHEREx would:
\begin{itemize}
\item Probe the origin of the Universe by constraining the physics of inflation, the exponential expansion of the Universe that took place a fraction of a second after the Big Bang, by measuring galaxy redshifts over a large cosmological volume;
\item Investigate the origin of water and biogenic molecules in the early phases of planetary system formation – from molecular clouds to young stellar systems with planet-forming disks – by measuring absorption spectra to determine the abundance and composition of ices;
\item Chart the origin and history of galaxy formation through a deep mapping survey.
\end{itemize}
Details concerning the SPHEREx mission and these three major science themes are available at \href{http://spherex.caltech.edu}{http://spherex.caltech.edu}, and SPHEREx’ core extragalactic and cosmological science is described in an earlier White Paper \cite{Dore:2014cca}.


As attractive as these three scientific questions are, they are far from encompassing the complete scientific grasp of SPHEREx.  As Table~\ref{tab:statistics} shows, SPHEREx’ unique all-sky database would include spectra of large numbers of astronomical and solar system targets, including data on extended and diffuse sources.  These data would enable a wide variety of scientific investigations, and the SPHEREx team is dedicated to making the SPHEREx data available to the scientific community to facilitate these investigations, which we refer to as Legacy Science. To that end, we have sponsored two workshops for the general scientific community to define the most interesting Legacy Science themes and to assure that the SPHEREx data products are responsive to their needs.  In February of 2016, some 50 scientists from all scientific fields met in Pasadena to develop these themes and the consequent implications for the SPHEREx mission.  The results of this initial workshop are reported in \citet{Dore:2016}.  The discussion at the 2016 workshop highlighted many synergies between SPHEREx Legacy Science and other contemporaneous astronomical missions, facilities, and databases.  Consequently, in January, 2018 we convened a similar workshop at the Center for Astrophysics in Cambridge to focus specifically on these synergies.  This white paper reports on and summarizes the results of the 2018 workshop.   

The astronomical community has never previously addressed the scientific opportunities presented by an all sky spectroscopic survey, particularly one which covers the critical wavelengths between 0.75 and 5 $\mu$m.  Thus the science questions addressed at the 2018 workshop and summarized below should be taken as examples of how the data from SPHEREx would enhance the scientific results from other programs (and vice versa) and are not intended to be all-inclusive.  We hope that by publicizing the results of the workshop and, subsequently, by keeping the science community informed about the progress of SPHEREx as we move forward, that we will stimulate the creation of many additional applications of the unique SPHEREx data.

\begin{table*}[htb!]
\begin{center}
\begin{tabular}{r l} 
\hline
\multicolumn{2}{c}{The SPHEREx All-Sky Survey}\\
\hline
$\bullet$ & $>$ 1 billion galaxy spectra\\
$\bullet$ & $>$ 100 million high-quality redshifts\\
$\bullet$ & $>$ 100 million stellar spectra\\
$\bullet$ & $>$ 100 thousand ice absorption spectra\\
$\bullet$ & $>$ 1 million quasar spectra\\
$\bullet$ & $>$ 10 thousand asteroid spectra\\
\hline
\end{tabular}
\caption{SPHEREx maps 1.3 trillion spectral-spatial elements over the entire sky in each of four independent surveys.}
\label{tab:statistics}
\end{center}
\end{table*}

\section{No Mission is an Island}

\subsection{SPHEREx in Context}

\begin{figure}[!th]
\centering
\includegraphics[width = 0.7\textwidth]{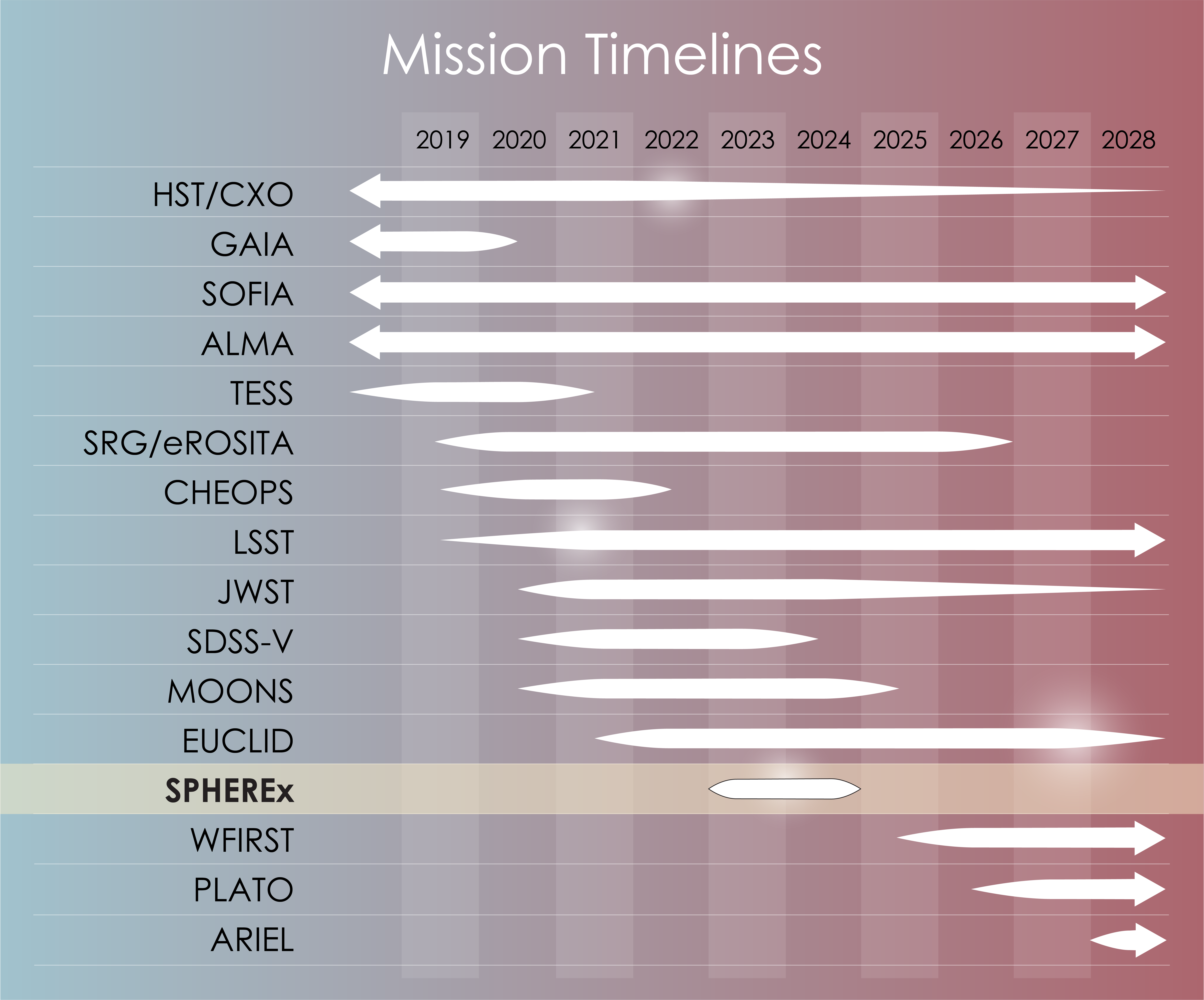}
\caption{Timeline of future space mission or major observatories in the 2020s.}
\label{fig:mission_timeline}
\end{figure}

\begin{figure}[!th]
\centering
\includegraphics[width = 0.7\textwidth]{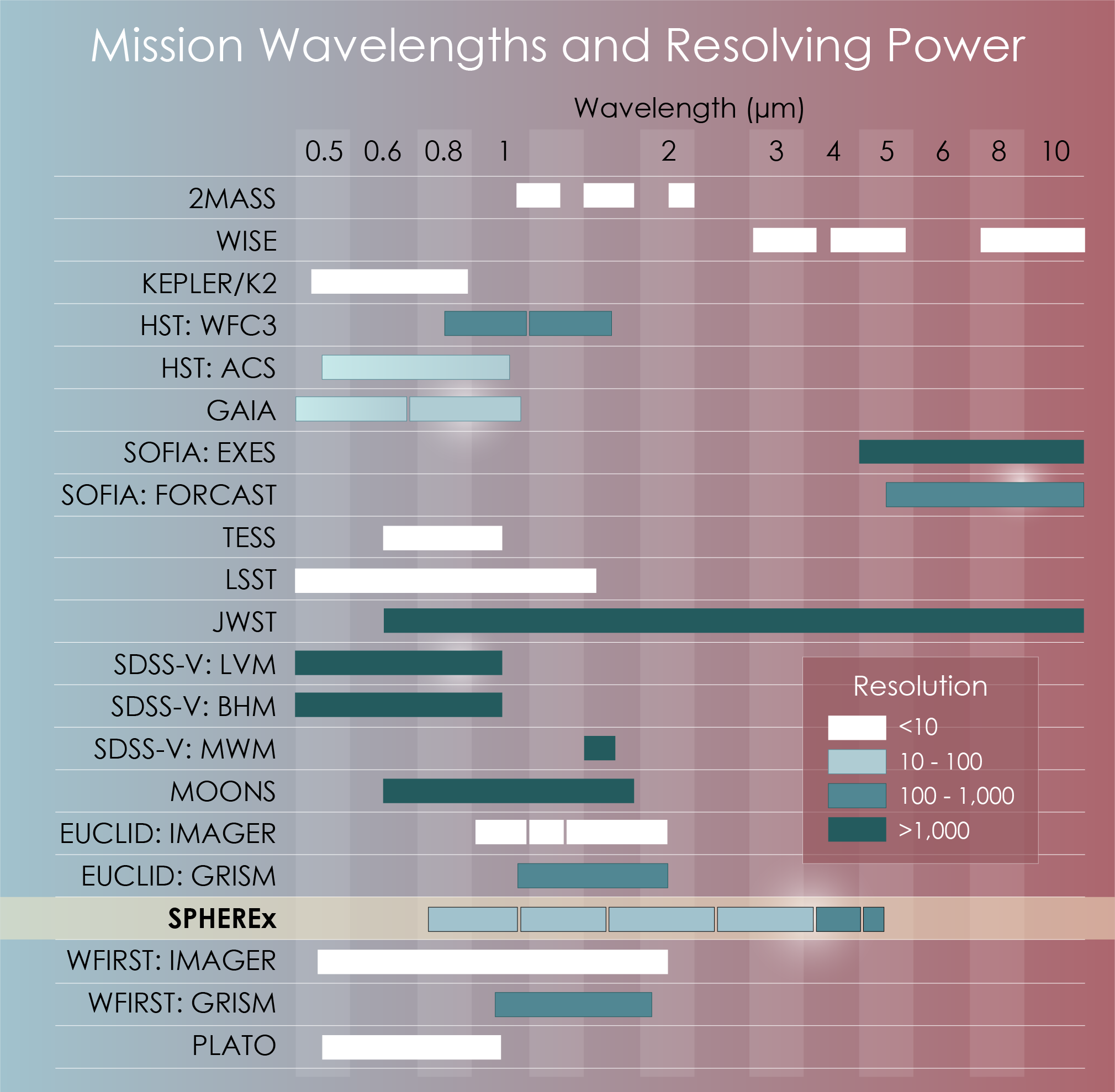}
\caption{Wavelength coverage and resolving power of future space mission or major observatories in the 2020s.}
\label{fig:mission_wavelength}
\end{figure}

The 2018 Workshop, and this summary, focused on synergies between SPHEREx and other major astronomical facilities or missions operating now and into the 2020’s.  Such synergies are widespread and of great importance, enhancing the scientific return from both SPHEREx and the other facilities.  Figures \ref{fig:mission_timeline} and \ref{fig:mission_wavelength} put SPHEREx in context with many of the other programs discussed in this white paper.  

The timeline in Fig.~\ref{fig:mission_timeline} shows that SPHEREx is well-timed to follow up on the results from missions such as JWST, TESS, and eROSITA; to identify targets for more detailed study by JWST, SOFIA, or ALMA, and to set the stage for later missions such as WFIRST and PLATO.  The wavelength/spectral resolution chart shown in Fig.~\ref{fig:mission_wavelength}  shows that SPHEREx provides almost unique access with significant spectral resolution to the wavelength range between 2.5 and 5 $\mu$m while extending shortward in wavelength to overlap with numerous ground-based and space-based imaging and spectroscopic surveys.  JWST is the only space-based mission with comparable wavelength grasp and spectral resolution, and this makes the synergy between SPHEREx and JWST particularly strong, as SPHEREx goes wide while JWST goes deep.  Not shown in either chart are the 10-30-meter class telescopes currently operating on the ground and planned for the future.  Their overlap with SPHEREx would be similar to that discussed below regarding JWST, with the additional consideration that they would not have full access even to the 2.5-to-5 $\mu$m wavelength region because of atmospheric absorption.  Nevertheless, we can anticipate that these telescopes, with their much higher spectral and spatial resolution, would be effectively used in pursuing targets and scientific questions identified by SPHEREx.

\begin{figure}[!th]
\centering
\includegraphics[width=0.7\textwidth]{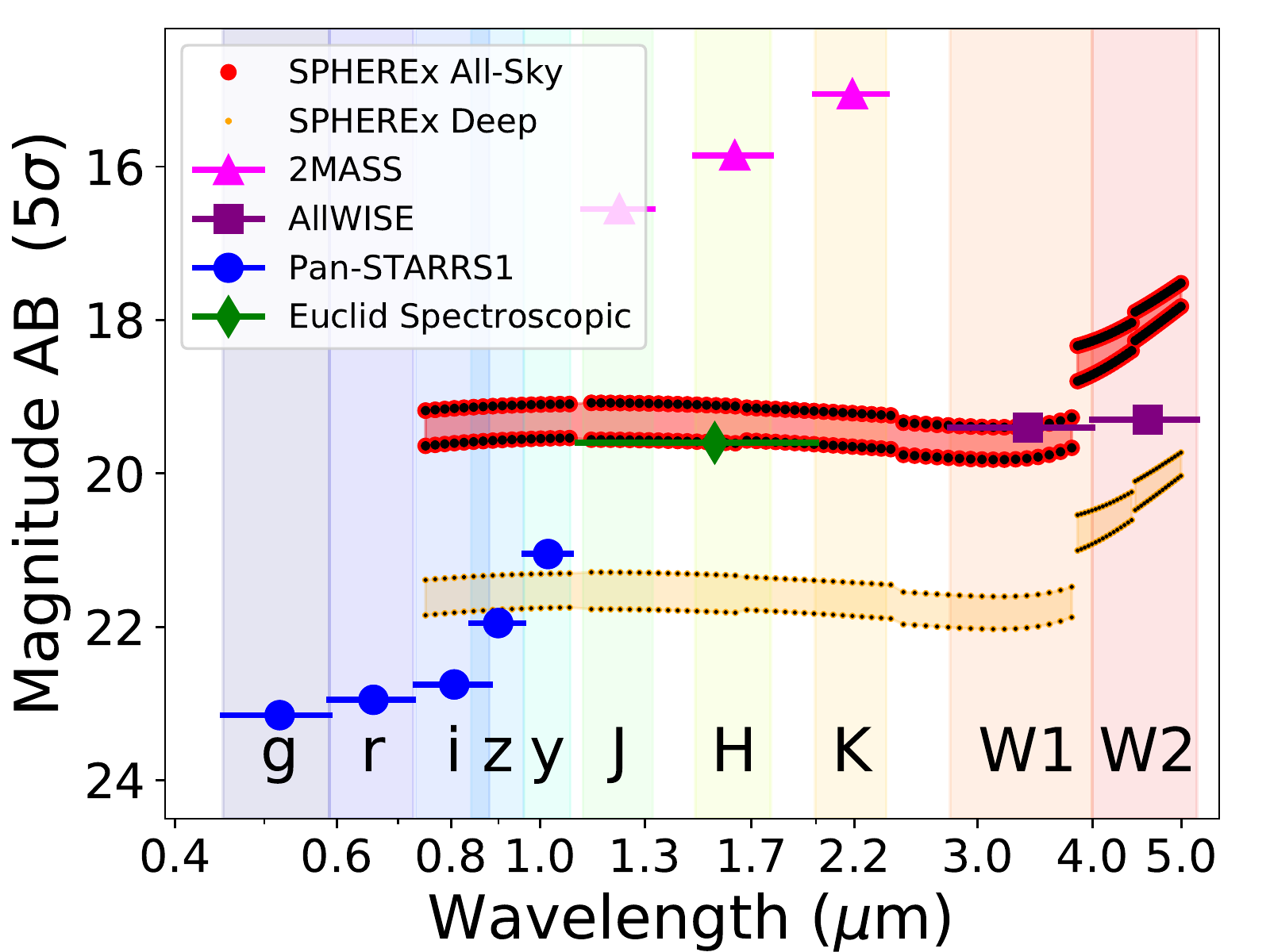}
\caption{Sensitivity of SPHEREx and current surveys (all at 5$\sigma$). The SPHEREX sensitivity is quoted for each $\lambda/\Delta\lambda$ = 41 spectral channel for 0.75 $<$ $\lambda$ $<$ 2.42 $\mu$m, for each $\lambda/\Delta\lambda$ = 35 spectral channel for 2.42 $<$ $\lambda$ $<$ 3.82 $\mu$m, for each $\lambda/\Delta\lambda$ = 110 spectral channel for 3.82 $<$ $\lambda$ $<$ 4.42 $\mu$m, and for each $\lambda/\Delta\lambda$ = 135 spectral channel for 4.42 $<$ $\lambda$ $<$ 5.00 $\mu$m. The lower of the two red curves corresponds to the current best estimate sensitivity over the whole sky, while the upper red curve corresponds to the instrument sensitivity based on specifications that each sub-system can meet with contingency over the  whole sky. The orange dots indicate the analogous sensitivity curves, but within the deep regions.  The statistical sensitivity does not include the effects of astrophysical source confusion, which is significant at the deep survey depth.}
\label{fig:lam_mag}
\end{figure}

While comparing SPHEREx capabilities with those of these other facilities it is of course important to have SPHEREx performance in mind.  This is shown in Fig.~\ref{fig:lam_mag}.

\subsection{SPHEREx and The Decade of Surveys}

All-sky surveys (e.g. IRAS, WMAP, Planck, GALEX, WISE) have played a major role in advancing modern astrophysics, enabling ground-breaking science and producing versatile legacy archives that have proven valuable for decades.  SPHEREx would contribute to this proud heritage by carrying out the first all-sky near-infrared spectroscopic survey.   The scientific value of a survey comes both from identifying rare objects of great interest and from the large data bases which support the archival research alluded to above.  As both the science and the instrumentation of modern astrophysics have evolved over recent years, astronomers are turning more and more often to surveys as tools for advancing our understanding of the Universe.  

It is perhaps not surprising, therefore, that of the 15 separate facilities and data bases listed in Figs.~\ref{fig:mission_timeline} and \ref{fig:mission_wavelength}, all but 4 (HST, JWST, SOFIA, and ALMA) refer to either unbiased or targeted surveys.  We shall see that the synergy of SPHEREx with these other survey missions is very strong, ranging from facilitating the identification of x-Ray sources imaged by eROSITA to refining the parameters of transiting exoplanets discovered by TESS by improving our knowledge of the exoplanet host stars.  Along the way, we will also note the important connection between SPHEREx’ extragalactic studies and those of the Euclid and WFIRST missions, to be launched in the 2020’s and featuring very large focal plane arrays optimized for studies of Dark Energy and other cosmological questions.  SPHEREx also plays well with ground-based surveys such as LSST and SDSS-V, overlapping the spectral coverage provided by these observatories and extending it out into the thermal infrared.  At the same time, we will see that SPHEREx is equally important as a partner of JWST and the other pointed observatories, both through singling out particularly interest targets for more detailed study and by working synergistically on numerous pressing scientific questions.

\section{2MASS and WISE}

2MASS and WISE are all-sky photometric surveys which overlap with much 
of the spectral range covered by SPHEREx. SPHEREx will obtain spectra of every sources in the 2MASS PSC (1.2 $\mu$m, 1.6 $\mu$m, 2.2 $\mu$m) catalog to at least (40 $\sigma$, 60 $\sigma$, 150 $\sigma$) per spectral channel, and spectra with S/N $\geq$3 per frequency element of the faintest sources detected by WISE (Fig.~\ref{fig:lam_mag}). SPHEREx thus adds a spectroscopic component to these widely used and scientifically flexible data bases, which have been used to study everything from the nearest cool stars and brown dwarfs to the stellar content of distant clusters of galaxies.

\section{JWST}

\begin{figure}[!th]
\centering
\includegraphics[width=0.98\textwidth,angle=0]{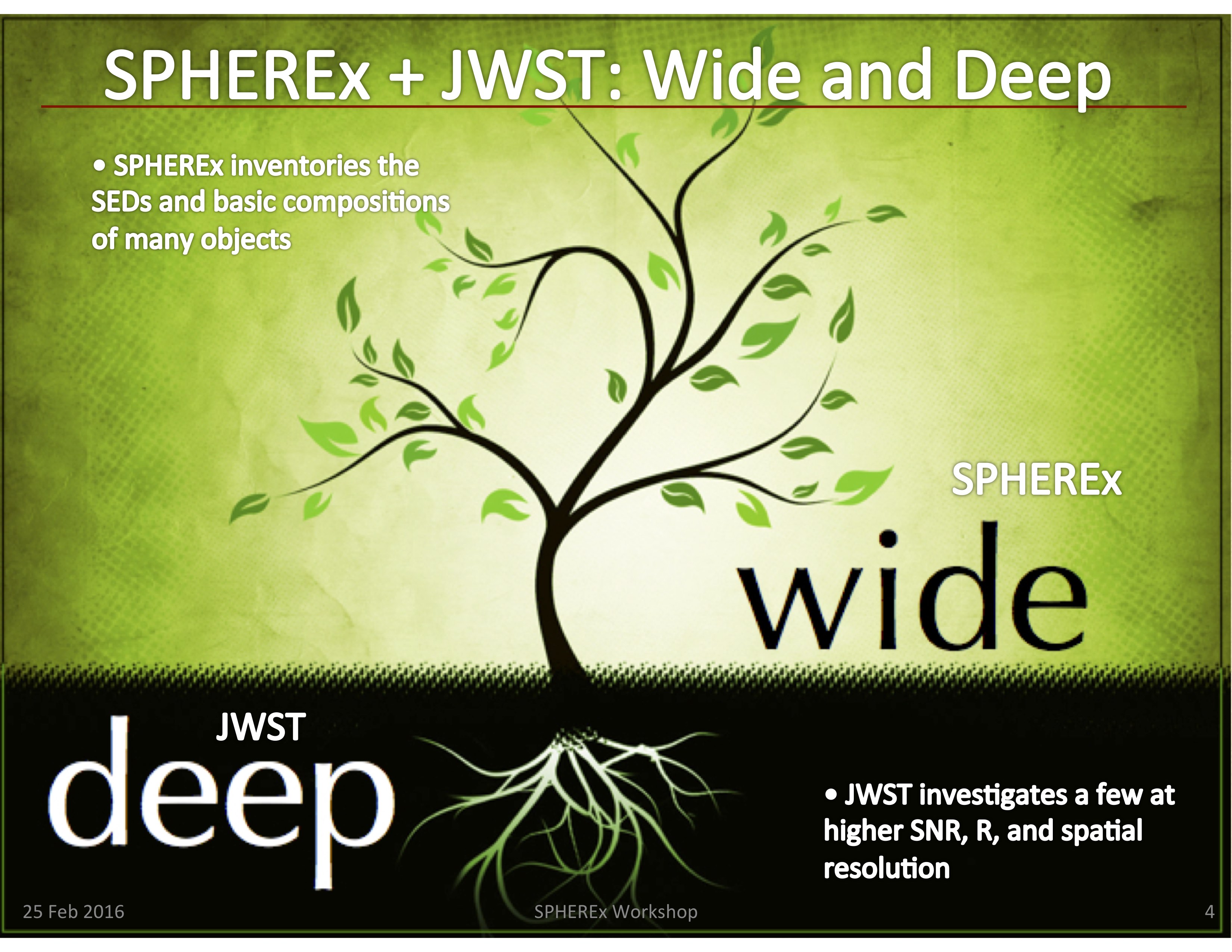}
\caption{Illustration of the synergies between SPHEREx and {\sl JWST}.}
\label{fig:tree}
\end{figure}

The SPHEREx all-sky surveys and legacy catalogs provide an invaluable archive of scientifically interesting sources for detailed followup and characterization with {\sl JWST} (Fig.~\ref{fig:tree}).

\subsection{Discovery Space}

As an all-sky survey, SPHEREx can identify rare and unusual objects for {\sl JWST.} A few examples include: (i) {\sl JWST} spectroscopy of rare $z >$ 6 (or 7) quasars discovered by SPHEREx, including high resolution imaging for their host galaxies; (ii) {\sl JWST} follow-up of rare planetary systems undergoing planetary aggregation or cataclysmic bombardment phase, identified by SPHEREx from their bright and variable near-infrared dust emission, and (iii) {\sl JWST} spectral imaging of objects with unusual ice spectra, suggesting stronger-than-expected isotopically shifted species or atypical elemental mixing ratios. The expected cadence of {\sl JWST} proposal calls and SPHEREx data releases should allow follow up of SPHEREx observations during the first year of the SPHEREx mission, 2023, and annually thereafter. All but the very brightest sources seen by SPHEREx will be observable by {\sl JWST} at R$\simeq$1000.

\subsection{Characterizing galaxies over time}

{\sl JWST} and SPHEREx can work together to study the evolution of galaxies over space and time. SPHEREx will begin by providing complete spatially resolved images of nearby galaxies, detecting infrared diagnostic lines such as CO, H$_2$O, H$_2$, HI, and PAH emission that probe the interplay of stellar populations with the Interstellar Medium (ISM). SPHEREx will also be sensitive to the evolved populations which constitute most of the stellar mass of these galaxies. The penetrating power of infrared observations will trace these phenomena into the dense, dusty, and highly active inner regions of galaxies.

Outside the local universe, {\sl JWST} images of galaxies selected on the basis of their SPHEREx spectra can be compared with those of the nearby galaxies imaged by SPHEREx, to investigate how the distribution of stars and the ISM have evolved over more than half the lifetime of the Universe. These studies, in turn, will set the stage for interpretation of {\sl JWST} observations of ever more distant galaxies, for which even this very powerful telescope obtains mainly the integrated properties. 

\subsection{M dwarfs and Brown Dwarfs}

\begin{figure} 
\centerline{\hbox{\includegraphics[width=6in]{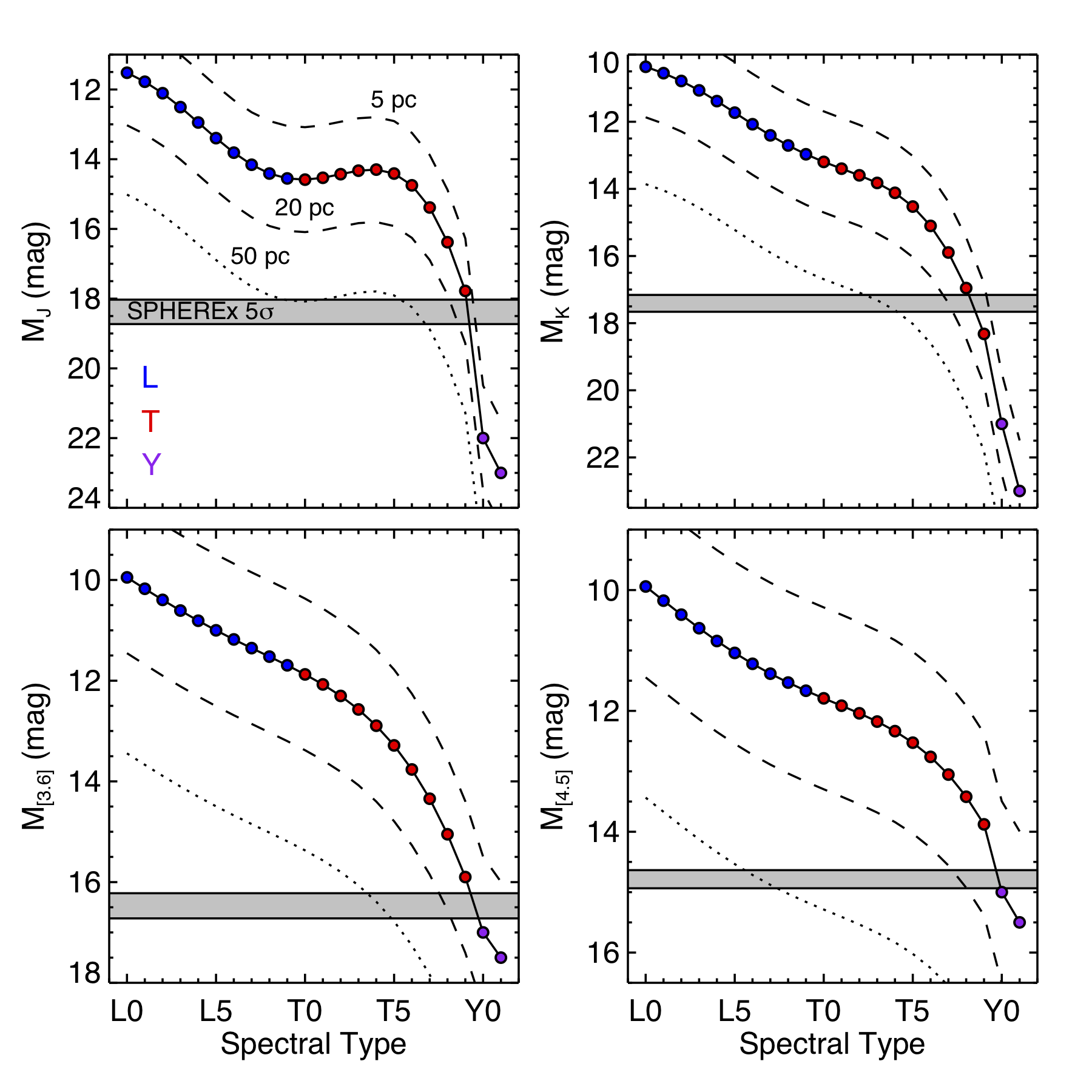}}}
\caption{Absolute magnitude versus spectral type for L, T, and Y dwarfs as a function of spectral type for four infrared wavelengths:  $J$, $K$, [3.6], and [4.5].  The data for the L and T dwarfs are from \citet{2012ApJS..201...19D} while the data for the Y dwarfs are from \citet{2015ApJ...799...37L,2017ApJ...842..118L}.  The 5$\sigma$ SPHEREx sensitivities are shown as grey bars and curves for distances of 5 and 20 pc (dashed), and 50 pc (dotted) are shown. At any wavelength, any object lying above the gray bars is detectable.}
\label{fig:brown_1}
\end{figure}

The SPHEREx wavelength coverage is nearly ideal for characterizing very low-mass stars and brown dwarfs (M, L, T, and Y dwarfs) because SPHEREx spectra span the thermal peaks of their spectral energy distributions and sample important absorption bands, e.g., H$_2$O, CO, CO$_2$, and CH$_4$.  Fig.~\ref{fig:brown_1} shows the absolute magnitude of known L, T, and Y dwarfs as a function of spectral type at four wavelengths:  
$J$ (1.25\,$\mu$m), $K$ (2.2\,$\mu$m), 3.6\,$\mu$m, and 4.5\,$\mu$m.  SPHEREx will obtain complete spectra from 0.7 to 5.0\,$\mu$m of many hundreds of L dwarfs and most of the known T dwarfs.  In particular, SPHEREx will drastically expand the number of L and T dwarfs with 2.5--5 $\mu$m spectra from of order tens of objects \citep[e.g.,][]{2000ApJ...541L..75N,2005ApJ...623.1115C,2009ApJ...702..154S} to nearly all such objects presently known.  (Spectral S/N ratios for the late-type T and Y dwarfs will be a strong function of distance.  Indeed the coolest brown dwarfs will only be detectable at the peak of the 5\,$\mu$m opacity hole). 

The 2.5--5\,$\mu$m spectral region is of particular interest because it contains bands arising from CH$_4$ \citep{2000ApJ...541L..75N}, CO \citep{1997ApJ...489L..87N}, and CO$_2$ \citep{2010ApJ...722..682Y}, the strengths of which are sensitive to the vigor of vertical mixing within an atmosphere.  Mixing within an atmosphere can dredge CO- and CO$_2$-rich gas from the hot, deep layers of an atmosphere to the cool, upper regions of an atmosphere where carbon is typically locked in CH$_4$.  The change in the band strengths of these molecules can then be used to measure the intensity of the mixing \citep{2003IAUS..211..345S}.  Previous work has been limited to a handful of spectra and \textit{Spitzer}/IRAC photometry \citep[e.g.][]{2007ApJ...655.1079L,2009ApJ...695..844G,2010ApJ...722..682Y}, so SPHEREx will revolutionize our understanding of vertical mixing in brown dwarf atmospheres.    

\begin{figure}[!th]
\centering
\includegraphics[width=0.95\textwidth,angle=0]{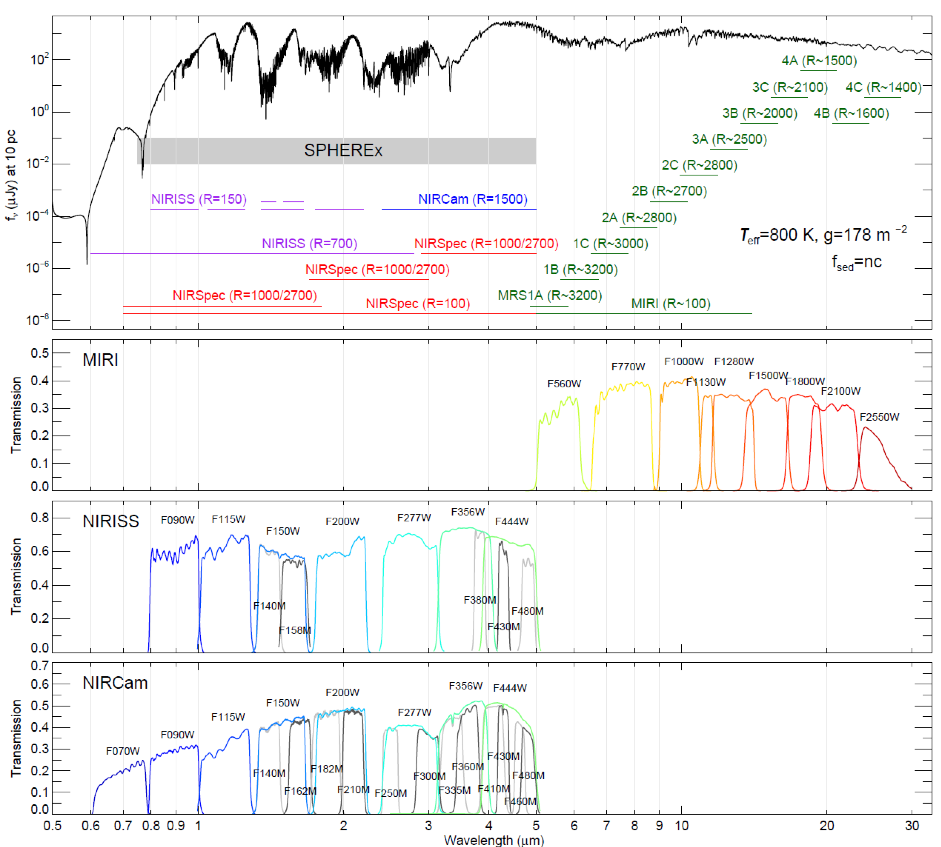}
\caption{Studies of brown dwarfs are one of the many areas in which \textit{JWST} can follow up on SPHEREx results, in this case by obtaining higher resolution spectra shortward of 5\,$\mu$m and/or extending the spectral coverage to longer wavelengths.  This figure provided by Mike Cushing compares the predicted spectrum of a late T dwarf with the spectral coverage of SPHEREx and that of the spectroscopic modes provided by \textit{JWST}.}
\label{fig:brown_dwarf_jwst}
\end{figure}

SPHEREx spectra combined with \textit{JWST}/MIRI spectroscopy and photometry will sample $>$80\% of the emergent flux of brown dwarfs having spectral types earlier than roughly T7 as visible in Fig.~\ref{fig:brown_dwarf_jwst}. SPHEREx$+$\textit{JWST} will therefore provide the most accurate measurements of $F_\text{bol}$ to date because the correction to account for the remaining flux will not dominate the error budget.  When combined with parallaxes from other missions or surveys such as Gaia, LSST, and \textit{Spitzer}, bolometric luminosities can be then computed.  Brown dwarfs obey a mass-luminosity-age relation and so for those brown dwarfs in nearby young moving groups such $\beta$ Pictoris ($\sim$15 pc), AB Doradus (20.1$\pm$1.6 pc), Tucana-Horologium ($\sim$48 pc), and TW Hydrae (53$\pm$2 pc) \citep{2016IAUS..314...21M}, $L_\mathrm{bol}$ can be used to infer their masses \citep[e.g.,][]{2017ApJS..228...18G}.  For those objects in the field that lack meaningful age estimates, we can still measure accurate effective temperatures using the Stefan-Boltzmann law $T_\mathrm{eff} = \left ( L_\mathrm{bol}/ 4\pi R^2 \sigma \right )^{1/4}$ by exploiting the fact that due to the equation of state, all brown dwarfs have nearly constant radii at $R\approx R_\text{Jup}$ \citep{2001RvMP...73..719B}.

\subsection{Dust Around Main Sequence Stars}

The collisional processes that creates planets leaves behind debris in the form of planetesimals (extra-solar analogs of asteroids, comets, Centaurs, and KBOs) and dust (zodiacal and Kuiper Belt dust).  The dust, heated by the star and replenished from the planetesimals, can frequently be observed, and studies of these debris disks has proven to be a powerful means of probing the nascent exoplanetary systems.  Until recently, debris disk studies have concentrated on wavelengths longward of 5\,$\mu$m because their contrast relative to the stellar photosphere increases markedly with wavelength.

Recent {\sl Spitzer} observations of extreme debris disks at 3.6 and 4.5\,$\mu$m have returned striking results which SPHEREx will build upon.  The brightness of debris disks simply reflects the amount of dust orbiting the star, so extreme disks (systems with $L_{dust}/L_{\star} \ge$ 0.01, about two orders of magnitude greater than what had previously considered to be bright debris disks) are unusual not only in having both a large amount of circumstellar dust, but also in having dust warm enough to radiate strongly shortward of 5\,$\mu$m.  Several such disks have now been identified orbiting stars younger than $\simeq$ 200\,Myr.  They have one other unusual property: their infrared radiation shows marked variability on time scales of months to years \cite{Meng:2014,Meng:2015}.

These extreme debris disks vary much more rapidly than predicted by evolutionary models in which the dust originates in a gradual collisional cascade. For the case of ID8 in NGC2547, for example, Meng et al.\ suggest  that the variability results from a major collision between two large 
(100--1000\,km radii) bodies of which some material might be vaporized, and then recondense rapidly into particles the size of sand grains.  Those particles would subsequently be degraded by collisions into the particles seen in the infrared, and blown out of the system.  Truly titanic collisions may be needed to produce the massive dust clouds required to account for the variability of the extreme debris disks.  Several such collisions are thought to have taken place in the first 100\,Myr of our Solar System, and others may have been triggered in tandem with the Late Heavy Bombardment which caused comets and asteroids to rain down on the terrestrial planets when the Solar System was 600--800\,Myr old.

New IRTF/SpeX studies (Lisse+ 2012, 2015, 2017a,b) have shown that not only can near-infrared 3-5 um spectroscopy detect thermal excess emission in more moderate debris disks, but that it can also detect systems which have been mistakenly classified as main sequence stars, and instead still have  remnants of their primordial birth-disks surrounding them. These systems, like their Class I and II YSO cousins, will be easily discerned by SPHEREx due to their "flat", very non-photospheric 1-3 um spectra and are in need of being thoroughly catalogued by an all-sky survey. 

SPHEREx can discover systems of any type in which the dust emission rises above the stellar photosphere at 3\,$\mu$m, and the six-month cadence of SPHEREx surveys is well-matched to the variability timescale of these extreme systems. It is important to note that today's well known debris disks were almost all found and identified using mid-infrared all-sky surveys; a proper unbiased,  senstivity limited near-infrared search for them has never been conducted. Particularly interesting candidates will be identified for {\sl JWST} follow-up, while SPHEREx refines our understanding of their prevalence and the attribution to the types of collisions described above. The {\sl JWST} observations can constrain the dust mineralogy, which may in turn place important constraints on the formation mechanisms.  The SPHEREx census will reveal the importance and frequency of YSO's, warm dust systems, extreme dust systems, and mis-typed stars, and help determine whether our own Solar System has a typical formation history, or if instead it stands out as an unusual system in the Milky Way.

As discussed further below, coordinated observations of circumstellar dust form an important portion of the synergies between SPHEREx and ALMA as well.

\subsection{Interstellar Ices}

Ices are an important yet relatively unexplored component of the interstellar medium and planet-forming disks.  Fewer than 250 ice absorption spectra have been obtained to date from the {\sl Infrared Space Observatory}, {\sl Spitzer}, and {\sl AKARI}.  Based on these limited spectra and gas-phase water data from the {\sl Submillimeter Wave Astronomy Satellite} and 
{\sl Herschel}, it is now clear that 99\% or even more of the water in dense clouds is locked in ice.  Along with organic molecules, ice-covered dust grains are predicted to be major repositories of the ingredients of life, and their incorporation into the larger bodies formed in protoplanetary disks may be a key element in how life can arise on habitable worlds.  Constraining ice column densities is key to modeling planet formation because the snow lines for different molecular species strongly affect the initial conditions for a model \cite{Qi:2013}.  SPHEREx's statistically significant study of ices across a broad range of source types, including the quantity (i.e., column density) of each ice species as well as chemical changes within the ice, will be a major contribution to our understanding of exoplanet formation and to exoplanet composition.

Many important biogenic ice features are accessible to SPHEREx by design, including those of H$_2$O (3.05\,$\mu$m), CO$_2$ (4.27\,$\mu$m), $^{13}$CO$_{2}$ 
(4.38\,$\mu$m), XCN (4.62\,$\mu$m), CO (4.67\,$\mu$m), and OCS (4.91\,$\mu$m).  During each of its four baseline all-sky surveys, SPHEREx is expected to obtain absorption spectra toward more than 10$^5$, and possibly as many as $2\times10^6$, Milky Way sources spanning all evolutionary stages from diffuse and dense clouds to young stellar objects and protoplanetary disks.  The diversity of objects observed by SPHEREx, coupled with the large number of sources for which high signal-to-noise spectra will be obtained, will constitute a dataset of unprecedented statistical power for ices in young objects.  This dataset, by measuring the relative amounts of these biologically important ices for so many systems over such a wide range of evolutionary stage down to newly forming planets, is expected to reveal whether the biogenically important ices form and evolve in situ, or conversely, whether these ices are delivered unaltered from their progenitor clouds.

{\sl JWST} can follow up interesting SPHEREx targets with higher resolution spectra over a broader wavelength range, picking up the ice features longward of 5\,$\mu$m.  The potential of JWST for ice studies will be demonstrated in an approved Early Release Science program to be carried out early in the mission; SPHEREx collaborator Karin \"Oberg is a member of {\sl JWST} ERS team 1309 (IceAge: Chemical Evolution of Ices during Star Formation; PI McClure) and will ensure good connections between the two missions in this important scientific area.	

\section{Euclid, WFIRST and LSST}

Euclid (launch 2021) and WFIRST (launch 2025) are visible and near infrared survey missions dedicated to studies of dark matter, dark energy, and related cosmological questions using a variety of observational probes over tens of thousands of square degrees. LSST, to be operational in 2022, will image more than half the sky at visible wavelengths. These observatories will enable a wealth of transformative astronomical studies. As an all-sky survey with instrumentation which is both overlapping and complementary, SPHEREx will enable new scientific opportunities in conjunction with Euclid, WFIRST and LSST.  As will be the case for JWST, the SPHEREx all sky surveys will identify many targets for the WFIRST GO program. 

\subsection{The Distribution of Galaxies in Space and Time}

The faint, distant sources seen by WFIRST and Euclid will be gravitationally lensed by the intervening galaxies seen by SPHEREx. This enables a cross-correlation measurement between these catalogs. The number of galaxies with high redshift accuracy provided by SPHEREx increases the significance of this signal by more than 50\%, allowing better studies of galaxy intrinsic alignment and formation \cite{Mandelbaum:2006,Mandelbaum:2013,Cacciato:2013}. Euclid and WFIRST spectroscopic surveys concentrate at redshifts z $>$ 1 to best study dark energy. By contrast, SPHEREx focuses at redshifts z $<$ 1. Combining these surveys will measure galaxy clustering over the full range over which dark energy begins to dominate the expansion of the Universe. 

SPHEREx enables new studies of galaxies which may intersect with Euclid and WFIRST in other ways. The SPHEREx catalog will contain $>$120 million galaxies with $\sigma(z)/(1+z) < 0.03$, sufficient for statistical galaxy evolution studies \cite{Ilbert:2013}, and enabling clustering studies that link galaxy properties to their underlying dark matter haloes \cite{Coupon:2012}.

\subsection{Overlapping Deep Fields}

\begin{figure}[!th]
\centering
\includegraphics[width=1.0\textwidth,angle=0]{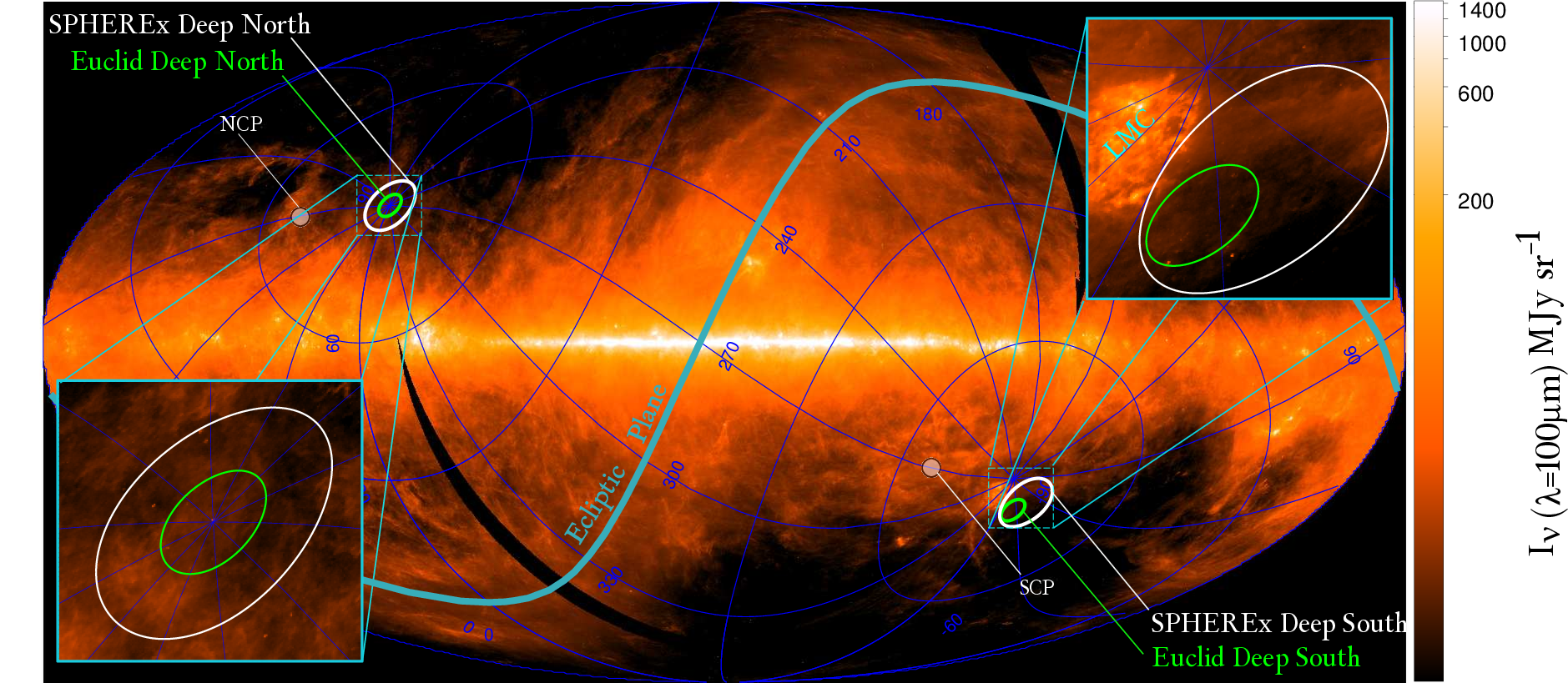}
\caption{SPHEREx deep fields at the ecliptic poles (white ellipses) overlap with LSST and Euclid deep fields (green ellipses).}
\label{fig:deep_field}
\end{figure}

SPHEREx orbital geometry naturally singles out $\sim$ 100 sq. degree regions near the ecliptic poles which will be surveyed to $\sim 10 \times$ the depth of the complementary all-sky survey.  Euclid, WFIRST, and LSST will all have deep survey regions of comparable angular extent.  As is shown in Fig.~\ref{fig:deep_field}, the Euclid and SPHEREx deep fields will overlap. SPHEREx deep fields will help to calibrate Euclid and WFIRST photometric redshifts.

\subsection{Synergy between SPHEREx \& Euclid: Focus on Photo-$z$ and galaxy evolution}

\subsubsection{Creating an Euclid+SPHEREx mock catalog}

We create a simple mock catalog by fitting \cite{Bruzual:2003} templates to the COSMOS2015 catalog from \citet{Laigle:2016}. We associate emission lines following recipes similar to \citet{Schaerer:2009}. This mock catalog covers 1.5 deg$^2$, with a detection performed in a $\chi^2$ image $zYJHK$ as deep as 24-25 mag in NIR (\cite{Laigle:2016}). We associate a spectrum (continuum+emission lines) to every source.

In a second step, we integrate these spectra into the expected transmission curves and we add noise to the predicted fluxes. Since our objective is an investigation of the combination of Euclid and SPHEREx, we integrate the spectra into the following filters:
\begin{itemize}
\item Euclid filter curves \cite{Laureijs:2011}, .i.e. a large band in visible noted $RIZ$ (24.5 mag at 10$\sigma$) and three NIR filters
  Y, J, H (24 mag at 5$\sigma$);
\item Optical filters with a possible configuration for the north
  hemisphere u, g, r, i, z (23.4, 24.5, 23.9, 23.6, 23 at 10$\sigma$);
\item 71 filters which mimic what SPHEREx would get between 0.75 and 4$\mu m$.
\end{itemize}
For SPHEREx, we adopte two configurations for the sensitivity,
considering a shallow survey at 19.5-19.7 and a deep survey at 22-22.2 mag. In this mock catalog, we ignore the difficulty of cross-matching the Euclid sources with their SPHEREx counterparts.

\subsubsection{Photometric redshifts}
\label{Sec:photoz}
We derive the photometric redshifts using the code Le$\_$Phare \cite{Arnouts:1999,Ilbert:2006}. We investigate three cases: Euclid alone, Euclid combined with SPHEREx shallow, Euclid combined with SPHEREx deep. We characterize the precision of the photometric redshifts using the Normalized Median Absolute Deviation
(NMAD) and we define as outliers sources with
$(z_p-z_{simulated}>0.15(1+z_{simulated}))$.  We also produce the statistic per bin of apparent magnitudes in Euclid $RIZ$ filter, which is the filter used to select galaxies for the weak lensing in Euclid. The resulting precisions are given in Table \ref{statistics}.

We first consider the Euclid+SPHEREx shallow case. As expected given the sensitivity of the SPHEREx data, SPHEREx will not be useful in improving the photo-$z$ for the general Euclid population at $RIZ<24.5$. However, for a population brighter than $RIZ<21$, SPHEREx will be able to produce extremely accurate redshifts. Moreover, such redshifts will be constrained by the presence of emission lines, which makes the
measurement less prone to biases in the photometry that could affect broad band surveys (see possible biases in \cite{Ilbert:2006}). SPHEREx will produce an excellent bright sample over the full Euclid area to
ensure that the quality of the Euclid photo-$z$ are homogeneous over the full survey. This reference sample will be complementary with the
Euclid NISP spec-$z$ which will be obtained for faint sources at $1<z<3$.  Moreover, the SPHEREx dataset will be extremely useful to isolate the AGN using the infrared power law \cite{Donley:2012}.

In the case of SPHEREx deep, the photo-$z$ are greatly improved, even at the magnitude limit of Euclid. At $23.5<RIZ<24$, we still expect an accuracy of 0.015. Such sample over 200 deg$^2$ will be extremely valuable to test the Euclid photo-$z$ at the faint limits, and eventually to train the Euclid photo-$z$ which are obtained with less bands.

\begin{table*}[htb!]
\begin{center}
\begin{tabular}{l c c c } \hline
                         &    Euclid      &     Euclid+spherex-shallow       &    Euclid+spherex-deep  \\ 
  RIZ  magnitude         &    NMAD        &     NMAD                         &       NMAD      \\ 
\hline
18-19	  &  0.013  &  	0.005	& 0.001    \\   
19-20	  &  0.012  &  	0.007	& 0.002    \\
20-21	  &  0.015  &  	0.010	& 0.004    \\
21-22	  &  0.020  &  	0.016	& 0.005    \\
22-23	  &  0.029  &  	0.025	& 0.007    \\
23-23.5	  &  0.040  &  	0.038	& 0.010    \\
23.5-24	  &  0.054  &  	0.052	& 0.015    \\
24-24.5	  &  0.078  &  	0.078	& 0.028    \\
 \\
\hline
\end{tabular}
\caption{ Precision expected by combining Euclid and SPHEREx per apparent magnitude bin into the $RIZ$ filter 
\label{statistics}}
\end{center}
\end{table*}

\subsubsection{Scientific complementarity for galaxy evolution}

By combining $H\alpha$ and $Pa\alpha$ measured by SPHEREx at $z<1$ and
$H\alpha$ measured by Euclid at $1<z<2.1$, we will obtain an excellent
picture of the instantaneous star formation rate over 80\% of the age
of the Universe. Note that $Pa\alpha$ is less affected by dust than
$H\alpha$ which will help in establishing dust correction.

\citet{Laigle:2018} showed that the reconstruction of the cosmic web in 3-D is feasible with photo-z having a precision $\delta_z/(1+z)$
better than 0.015. Therefore, the cosmic web can be reconstructed over the full sky using SPHEREx at $RIZ<22$. Galaxy evolution could be studied as a function of their position in the cosmic web. Using the exquisite morphology from the Euclid/VIS instrument, we could study
how environment impact the galaxy morphological transformation. Having simultaneously the SFR, we could link this transformation with the
quenching of the star formation.

Numerous other studies could benefit from combining SPHEREx and Euclid. Among many, lensing from Euclid, combined with the clustering and abundance analysis from SPHEREx at $z<1$ can be used to study the galaxy-halos connection (e.g. \cite{Leauthaud:2012,Coupon:2015}), and the 3-D cosmic web at $z<1$ could be compared with shear map to link baryonic matter and dark matter.

\subsection{Cosmology with Cosmic Voids}

When considering void catalog based cosmology, SPHEREx is unique when comparing to Euclid and WFIRST in providing high galaxy number density and all-sky coverage at lower redshifts (z$<$0.6). This provides an important anchor for probes of the expansion history and growth of cosmic structure during the acceleration epoch (the epoch of dark energy domination). While SPHEREx will add to the existing baryon acoustic oscillation constraints, its unique qualities make it a stand-out mission for probes of smaller scale structure. 

\begin{figure}[!th]
\centering
\includegraphics[width=0.7\textwidth,angle=0]{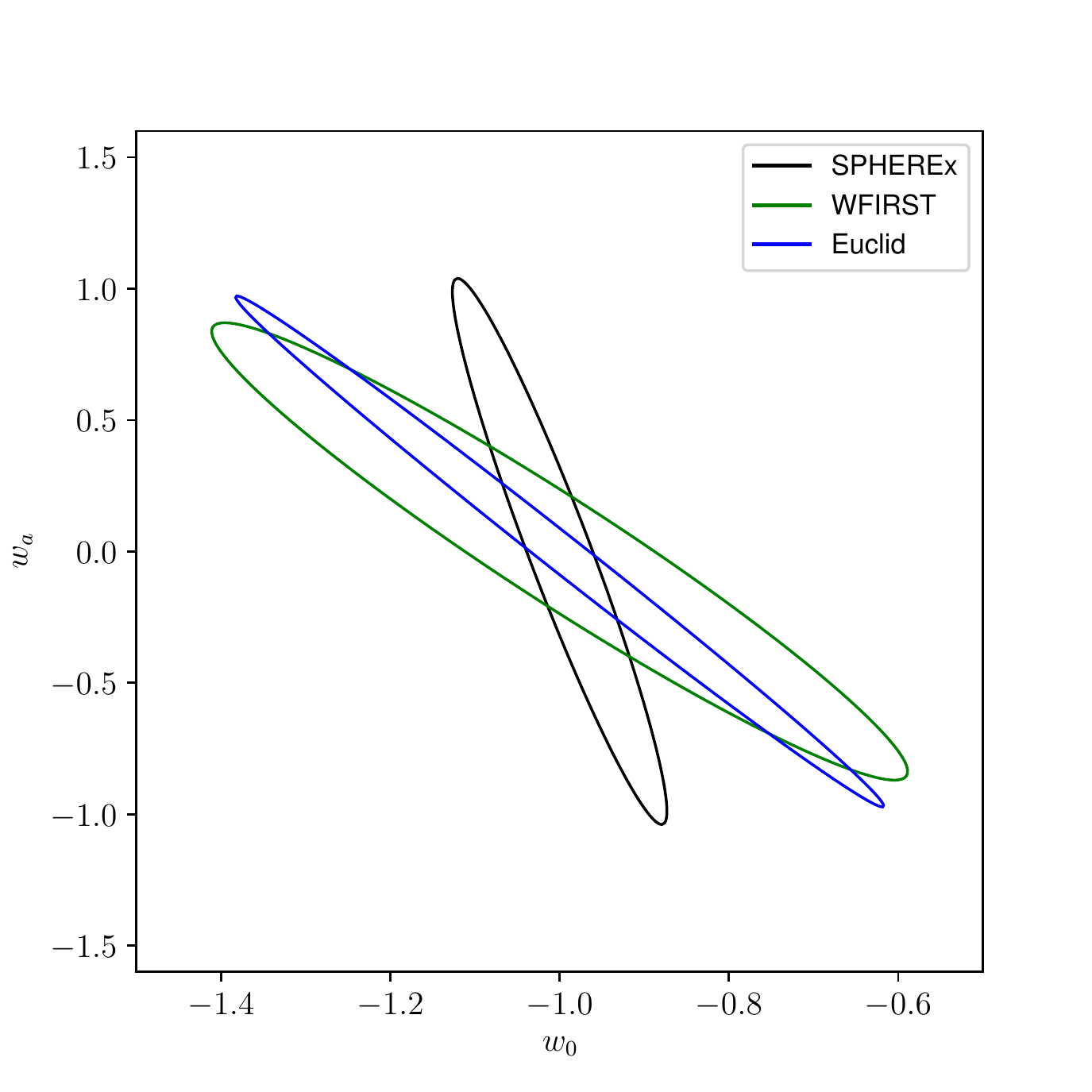}
\caption{Forecast constraints from void number counts on dark energy equation of state parameters (see \cite{Pisani:2015} for details). The complementarity of the constraint is a consequence of unique combination of redshift and sky coverage of the SPHEREx survey.}
\label{fig:voids_1}
\end{figure}

Several studies have shown (e.g. \cite{Hamaus:2016} and references therein) that voids can be modeled sufficiently precisely even on mildly non-linear scales (void radii of ~10 Mpc/$h$) to be used as a probe of the growth of cosmic structure and cosmography through the Alcock-Paczyinski test. While the Euclid and WFIRST missions are forecast to detect tens of thousands of voids around redshifts of $\simeq$ 1 and above, we predict that SPHEREx will find over 200,000 voids at lower redshift, probing the recent acceleration  of the universe. 
Other cosmological tests based on the number and size distribution of the detected voids \cite{Pisani:2015} are forecast to provide complementary constraints on the dark energy equation of state parameters to those from Euclid and WFIRST as shown in Fig.~\ref{fig:voids_1}. The large added value of SPHEREx compared to these much larger missions is a consequence of its unique redshift range and sky coverage.

\subsection{Exploiting Euclid-SPHEREx complementarity for primordial non-Gaussianity constraints}

An enticing science goal of the SPHEREx mission is the characterization of primordial non-Gaussianity through its imprint on the very large scale galaxy 2-point correlations. In terms of statistical power this measurement promises to improve by a significant factor the current best constraints derived from the three-point function of the cosmic microwave background anisotropy. This measurement will require an exquisite control of systematics on the largest scales, in particular systematics that modulate the observed fluctuation power. While the selection in an all-sky spectroscopic survey is more straightforward than in traditional redshift surveys, we still expect some sources of spurious clustering will be present. Detailed simulations will be key to assess the impact of these potential systematic errors on non-Gaussianity science with SPHEREx. While SPHEREx holds substantial margin against these systematic effects, again, the complementary survey strategies of SPHEREx and Euclid suggest promising approaches to characterize and mitigate these effects. For example, Euclid’s higher resolution will more clearly distinguish between stars and galaxies; coupled with SPHEREx's all-sky spectroscopic approach, a Euclid-selected SPHEREx sample may offer an optimal combination of systematics mitigation and statistical power. Reversely, SPHEREx calibration stability provides a linked network of infrared spectrophotometric standards over a wide range of fluxes and the entire sky.  We anticipate this network will be widely used across astronomy, benefiting current and future observatories, much like Gaia setting a new standard for calibration at optical wavelengths. It should be of great value to Euclid and help mitigate calibration drifts.




\subsection{Comets and Asteroids}

\subsubsection{Executive Summary}

SPHEREx all-sky survey images will contain NIR spectral detections of tens of thousands of asteroids, greatly increasing our ability to sort asteroids into evolutionary families and define compositional trends with heliocentric radii created by structure in the Solar System’s protoplanetary disk. SPHEREx will also be sensitive to emission from comets, the end reservoir for the interstellar ices studied under SPHEREx’s main themes. Of particular interest will be measurements of the relative intensities of CO and CO$_2$, both with emission bands in SPHEREx Bands 5 and 6. WISE and Spitzer have shown that these molecules can drive cometary activity at large heliocentric distances. SPHEREx should be able to measure these molecules in upwards of 100 comets. Both the asteroid and the cometary studies will be among the many ways in which SPHEREx will be synergistic with the survey results from the Large Synoptic Survey Telescope [LSST], which will survey the entire sky visible from Chile every three days at wavelengths from 0.35 to 0.95 $\mu$m. LSST images will be particularly useful in providing an estimate and image of cometary activity complementary to that inferred from the SPHEREx spectra. In turn, the SPHEREx + LSST measurements will be invaluable for selecting high value added ToO targets for HST, JWST, WFIRST, etc.

\subsubsection{SPHEREx Asteroid Spectral survey}

A major question in solar system formation science today is how and where the current population of asteroids formed. In the last decade the formation of iron meteorites, remnants of planetesimal cores, have been dated to within 1 to 3 Myr after meteoritic CAI's ("Calcium Aluminum Inclusions", mineral bits composed of the stablest and most refractory metal oxides that form first out of a cooling solar abundance mixture) formed the oldest known materials in the solar system. This has been interpreted as strong evidence for a "top-down" formation of large Vesta-sized asteroidal bodies in the innermost regions of the solar system, which then underwent collisional disruption over the next 10-30 Myr to produce collections of metallic (from the core), stony-iron (from the mantle), and stony (from the crust) fragments that re-accreted into the asteroid families we see today. Many of these bodies were absorbed in the making of the terrestrial planets. Those that weren't underwent further collisional grinding, to form the smaller bodies of the asteroid families we know today. Recent spectral survey work of $\sim$ 300 asteroids at 0.8 - 2.5 $\mu$m \cite{Demeo:2014} has shown that the asteroid belt also appears to be zoned by distance from the Sun, with the largest bodies keeping the signatures of their formation location: rockiest in the inner belt, mixed in the center regions, and most carbon and water-rich in the outer regions. Brand new iron meteorite isotopic evidence \cite{Kruijer:2017} is now arguing for 2 distinct reservoirs of solar system material separated by Jupiter’s core forming within 0.6 Myr of the CAIs.

WISE performed an all-sky asteroidal survey from 2010 - 2015, detecting $\sim$ 200,000 asteroids and determining their size, albedo, and color frequency distributions. SPHEREx will be able to spectrally characterize a large number of the WISE asteroids and determine their makeup as a function of orbital location, bridging the gap between the WISE and the DeMeo surveys \cite{Demeo:2014}.

SPHEREx will need to perform this asteroid survey for another reason - given the NIR brightness frequency distribution of the asteroid population and its extent across the sky, there is a few \% chance in a SPHEREx observation of any given sky pixel that there will be a significant asteroidal contribution to its measured flux. To make robust measurements of extra-solar system objects, the SPHEREx team will thus have to remove any foreground asteroid contribution, naturally building up an asteroid spectral survey. Fortunately, at SPHEREx sensitivity levels most of the asteroids detected will be previously cataloged objects and asteroids would appear as unusual rogue, variable sources in SPHEREx' redundant surveys.

The SPHEREX extended asteroid survey has direct NASA mission support consequences. SPHEREx will return the first large scale spectral characterization of the solar system’s Main Belt Asteroids (MBAs). As such, it will place the unique mission targets of past asteroid rendezvous missions  - like NEAR, Hyabusa, and DAWN – into better context vs. the whole MBA population, while doing the same for the new OSIRIS-ReX, and PSYCHE missions. E.g., it will help answer the question of how common a near-Earth C-type asteroid like the O-Rex target Bennu is, and if Psyche has many other little brother and sister M-type asteroids in the Main Belt, or is as nearly unique as it seems.

Even though SPHEREx can easily achieve WISE-like sensitivity levels on a fixed target, asteroids may pose challenges because of their motion and rotation in combination with the piecewise way in which SPHEREx will build up a spectrum. Thus we have conservatively estimated that SPHEREx will return useful data on tens of thousands, and not $\sim$200 thousand asteroids, as WISE has done. Even so, robust spectral characterization from 0.75-to-5.0 $\mu$m of tens of thousands of asteroids will be a major scientific advance over the $\sim$500 asteroids spectrally characterized in the NIR over the last 3 decades.


\begin{figure}[!th]
\centering
\includegraphics[width=0.75\textwidth,angle=0]{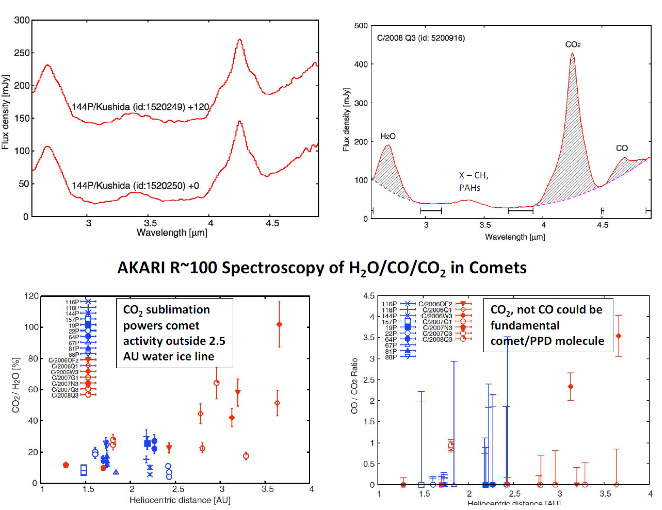}
\caption{(Top) Examples of the expected quality of cometary spectra from SPHEREx. Shown are AKARI 1-5 $\mu$m R=50 spectra of 12 comets (after \cite{Ootsubo:2012}). The areas under each spectral feature are directly related to the total amount of the emitting species in the AKARI beam. The main features evinced are due to the O-H stretch in water and hydroxyl (2.4 - 2.8 $\mu$m), the aliphatic and aromatic C-H stretch in organics like CH4, C2H6, H3COH, H2CO (methane, ethane methanol, and formaldehyde; 3.2 - 3.6 $\mu$m), the C=O stretch in CO2 (carbon dioxide; the doublet from 4.0 - 4.5 $\mu$m centered around 4.25 um), and the C=O stretch in CO at $\sim$ 4.7 $\mu$m (carbon monoxide). (Bottom) Trends in the estimated CO$_2$/H$_2$O and CO/CO$_2$ ratios. A clear trend of rising CO$_2$ vs H$_2$O production is seen as the observed comets move outside the water ice line at $\sim$ 2.5 AU and water production from the nucleus shuts down. In contrast, trends in the CO$_2$/CO ratio are hard to distinguish because both species are still highly volatile at a few AU while not being trapped in solid water ice. Like AKARI, SPHEREx will remove the confusion in previous 4.5 $\mu$m photometric surveys (e.g., WISE) between flux produced by CO2 vs CO in the 4-5 $\mu$m range. But improving on AKARI, it will do this for $>$ 100 comets, and at much higher sensitivity due the greatly reduced thermal instrumental background because of its much colder optics bench.}
\label{fig:lsst_comets_1}
\end{figure}

\subsubsection{Comet Chemical Abundance Survey}

Comets, formed within the first Myr of the solar system's lifetime, are thought to be the most primitive bodies left over from the Proto-Planetary Disk (PPD) era. They are leftover relics from the era of planetary formation that have failed to aggregate into a planet (or looked at another way, survived the era of violent early aggregation). There are two main reservoirs of comets known today in the solar system: the Kuiper Belt and the Oort Cloud. The Kuiper Belt comets are small icy bodies that formed at the outer edges of the PPD, where there was too little mass density to form another planet; looked at another way, the PPD had to end somewhere, and in its regions of lowest density planetesimals accretion was slow and truncated. 

These objects are in orbits relatively well clustered around the plane of the ecliptic with low inclinations (modulo the objects scattered by Neptune as it migrated outward during the LHB), and exist in a radial zone extending out to $\sim$ 50 AU from the Sun. We observe them when they become scattered inward into the giant planet region by galactic tides and passing stars, becoming first Centaurs and then finally short period comets. (The large outer planet moons Phoebe and Triton and comet/Centaur 29P/SW1 are direct examples of how scattering can send relic icy bodies inward.) The Oort Cloud comets, found in a roughly spherical distribution from 103 to 105 AU, were counterintuitively formed inside the Kuiper Belt in the giant planet region, as the feedstock for the nascent giant planets; they represent the population of objects that had near misses to the growing giant oligarchs in the first 1-10 Myr of the solar system, and rather than accreting or being thrown into the Sun or out of the system entirely, they were scattered into highly elongated, barely bound, Myr orbits. One of the holy grails of comet science over the last 2 decades has been to search for compositional differences between the Kuiper Belt and Oort Cloud comets, as a signature of radial chemical gradation in the PPD. SPHEREx will observe large enough numbers of comets to start to fill this data gap, particularly if more than the expected number of Oort Cloud comets appear during its 2 year mission.

The search for PPD chemical signatures in comets has produced another important result. Cometary bodies are composed of $\sim$ 1/2 icy volatiles and 1/2 rocky refractory materials, with the ices being dominated ($>$ 80\%) by H$_2$O ice. The most important ice species after water are CO and CO$_2$ \cite{Bockel:2004}, with minor admixtures of methane, ethane, formaldehyde, methanol, and ammonia. Until 2012, when Ootsubo et al. used the Akari satellite to observe 18 comets from 1-5 $\mu$m at R$\sim$50, it had long been thought that CO was the fundamental C-bearing icy reservoir; but we now know that while CO$_2$ ranges from 5 to 25\% vs. water in abundance in comets, CO can vary over 3 order of magnitude in relative abundance, from from 0.1 to 30\%. With the recent discovery of abundant O$_2$ in comets by ROSETTA \cite{Bieler:2015,Keeney:2017}, the higher stability of the more oxidized CO$_2$ form becomes more plausible, as does the variability in the likely original source CO (c.f. interstellar ices section).) Thus CO$_2$ may be the true leader of the C-bearing family.

Much work has been done in the last 5 years to quantify the amount of C-bearing gas in comets. As emission lines from these species are best detected from space (and CO$_2$ is detectable only from space) owing to the Earth's atmospheric absorption, space-based platforms are the best-equipped facilities for characterizing their production by comets. WISE and Spitzer have undertaken photometric campaigns to identify CO and CO$_2$ emission in comets, and to characterize their production as a function of distance from the Sun and of comet orbital class. However, unlike AKARI’s spectral cometary measurements, it was not possible for these spacecraft to directly differentiate between CO and CO$_2$ production, as both molecules’ main emission features lie in the same broadband 4.5 $\mu$m filter. 

Furthermore, regarding comparison to water production, because of the variable nature of comets and the lack of the ability of these space platforms to quantify water production simultaneously, placing the combined production limits of CO and CO$_2$ in relation to the most plentiful species produced in the inner solar system by comets is not certain. Even though the water production may be characterized from the ground, the ground-based observations are rarely simultaneous. SPHEREx will be able to detect these C-bearing species separately, and simultaneously with water and other astrobiologically important volatiles such as the organics methane, ethane, methanol, formaldehyde, and ammonia.

The spectral imaging capabilities of the spacecraft will play a special role in more detailed investigations of particular species. Comets are extended objects containing a compact nucleus emitting dust and gas into an extended coma and tail. By measuring the column density profiles of a species as a function of its distance from the comet nucleus, they will facilitate the accurate measure of its solar UV photolytic dissociative scale length, a feature of cometary coma chemistry which still remains ambiguous for many important species (like CO \& CO$_2$), and so help investigate the physical processes (like radiative photolysis, charge transfer, solar wind particle collisions, and gas phase and dust surface reaction chemistry) occurring in a comet’s coma. By measuring the continuum color structure in the extended dust coma and tail, we will be able to use SPHEREx observations to determine the size distribution of dust emitted by the comet and size sorting effects on the dust as it flows away from the comet under the influence of solar radiation pressure and gravity, and thus estimates of the comet’s total emitted dust mass and dust mass emission rate.

In summary, by performing an unbiased spectral survey of  $\sim$ 100 comets (the expected number based on the comets detected in 2 years by \cite{Bauer:2015}. SPHEREx will be able to directly expand our knowledge of the quantity of dust, organics, CO, and CO$_2$ in comets relative to H$_2$O, and so provide compositional constraints on the part of the PPD in which they formed. SPHEREx will further provide independent estimates or constraints on the size, albedo, and activity level of these comets, as a function of time and heliocentric distance (- local equilibrium temperature). Finally, there is the potential for enormous synergy with contemporaneous small body surveys, such as NEOCam, which produce broad-band mid-infrared detections of hundreds of thousands of potentially hazardous impactors, both cometary and asteroidal, but without much compositional information about them other than their absolute albedo (dark = carbon rich, bright = icy, middling = stony or iron). SPHEREx will not only improve the albedo estimates with its own independent albedo measures, but it will also provide the spectral measurements needed to distinguish in detail a cometary vs a stony vs a metallic impactor, which is vital for implementing accurate mitigation schemes for any impact potential threat. 
 
\begin{figure}[!th]
\centering
\includegraphics[width=0.75\textwidth,angle=90]{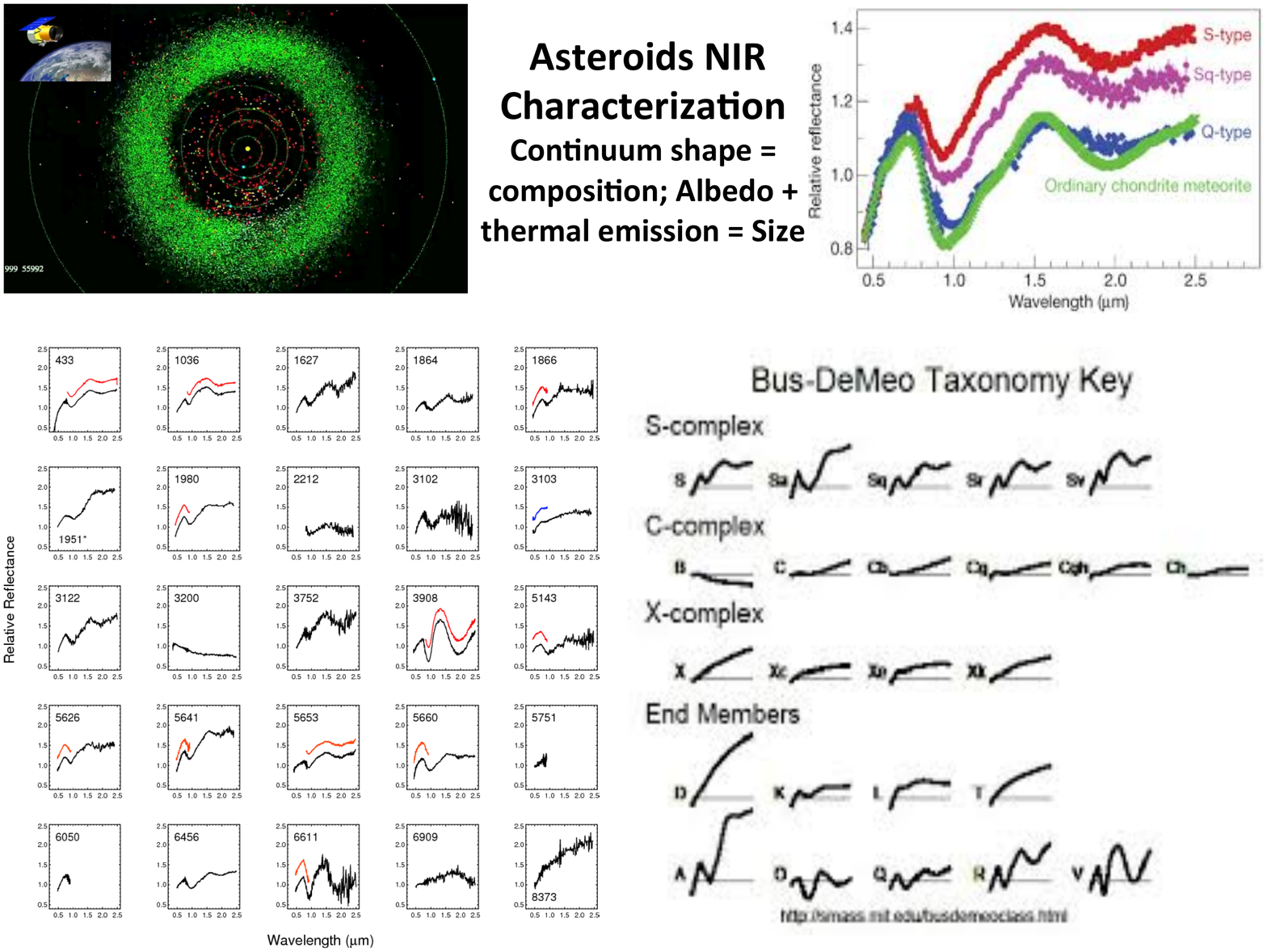}
\caption{Illustration of asteroids compositional classification using near-infrared reflectance spectroscopy.
Shown, starting in the upper left and moving counterclockwise are: the WISE spatial distribution of main belt 
asteroids (MBAs); 0.8 - 2.5 $\mu$m spectra for 25 asteroids taken from the NASA/IRTF 3m on top of Mauna Kea; 
\cite{Demeo:2014} spectral templates for different classes of asteroids, referenced to laboratory measurements 
of meteorite samples; and in the upper right, a detailed comparison of 3 closely related rocky asteroids, showing 
the subtle differences in the olivine and pyroxene absorption features that sets them apart. The dominant 
distinguishing spectral reflectance features observed are the 1.2 um and 1.9 $\mu$m olivine absorptions, the 1.3 $\mu$m 
pyroxene absorption, and the broad shallow 0.8 - 2.5 $\mu$m continuum reddening due to carbonaceous material.\label{fig:lisse5}}
\end{figure}
 
\section{LUCY and the Trojan \& Greek Asteroid Survey}

\begin{figure}[!th]
\centering
\includegraphics[width=0.375\textwidth,angle=0]{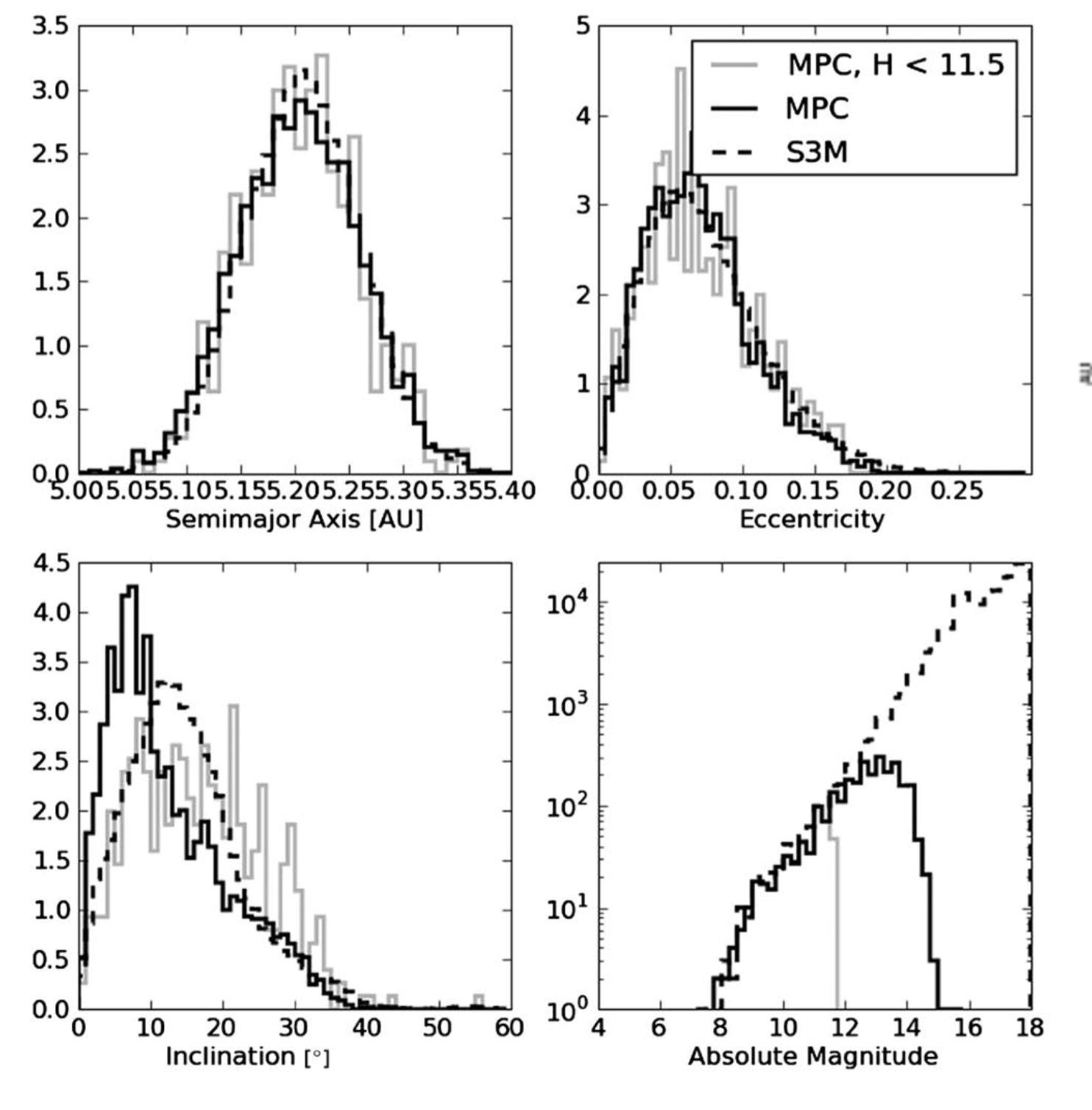}
\includegraphics[width=0.375\textwidth,angle=0]{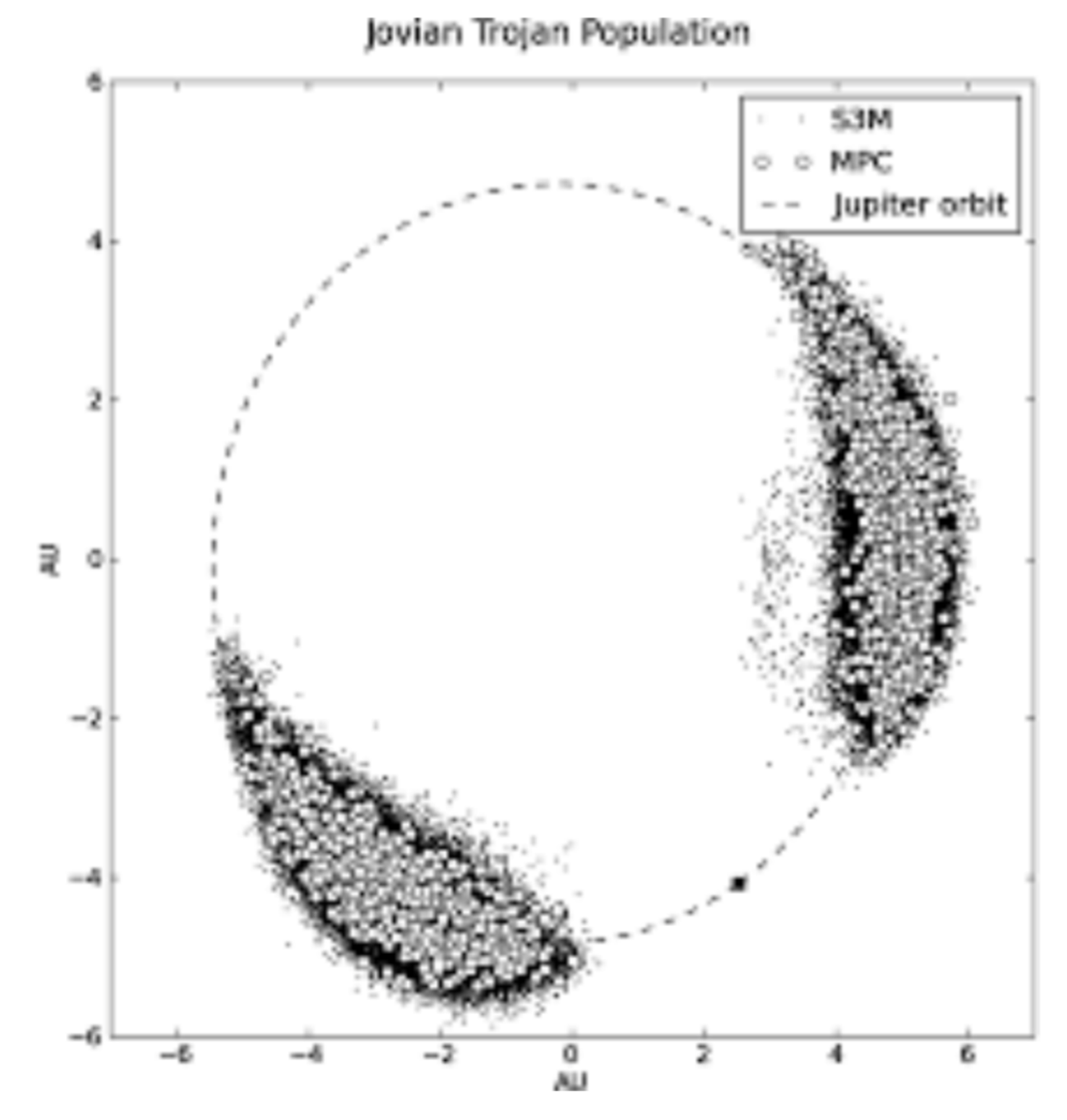}
\caption{\emph{Left:} Estimated brightness and orbital frequency distributions for the Trojan and Greek asteroids, after \cite{Grav:2010}. It is estimated that there are millions of slow moving bodies in the clouds that SPHEREx could survey. The vertical grey line at absolute magnitude $\sim$ 12 denotes the predicted WISE/SPHEREx sensitivity limit, so there should be hundreds of known Trojan asteroids that SPHEREx will be able to characterize spectrally for the first time during its 2-year mission, compared to the handful that have been observed to date. \emph{Right}: Spatial location of the Greek asteroids leading Jupiter in its orbit around the L4 resonance ($\sim$ 60 deg. in the prograde direction) and of the Trojans trailing Jupiter in its orbit around the L5 resonance ($\sim$ 60 deg. in the retrograde direction). Jupiter is the small dot at (2.3, -4.2) AU. Note that the Trojans and Greeks are safely removed from Jupiter and each other, so that there is no likelihood of scattered Jovian light causing problems with the SPHEREx measurements, while at the same time it will require a dedicated survey like SPHEREx to sample the spatially extended swarms well. LUCY will pass through both the Greek and Trojan swarms. Distributions of semi-major axis (top left), eccentricity (top right), inclination (bottom left), and absolute magnitude (bottom right) of the synthetic (dotted line) and known (black solid line) populations of the Jovian Trojan population. The distributions of the sample of known objects with H$<$ 11.5, a sample believed to be $\sim$ 90\% (gray solid line).}
\label{fig:trojans_1}
\end{figure}

In a similar fashion, SPHEREx will naturally observe and detect asteroidal bodies located in the L4 and L5 resonances of Jupiter, the so-called Greek and Trojan asteroids also known as "Jupiter Trojans". Characterizing these objects will be a natural offshoot of the main belt asteroid survey. Even though their total number is predicted to exceed the number of MBA's, only a fraction of these objects have been studied spectroscopically in the NIR. \cite{Grav:2010} studied 1742 Jupiter Trojan Asteroids using the WISE spacecraft broad-band photometry, and found 3.4 $\mu$m albedo differences between C \& P-spectral-type and D-types. SPHEREx will provide a better understanding of the compositional causes of this color correlation. SPHEREx should be able to obtain good spectra of hundreds of Trojan and Greek asteroids; they move across the sky more slowly and predictably than the main belt asteroids and should present fewer observational problems (Fig.~\ref{fig:trojans_1}).

The importance of SPHEREx information about the composition of these bodies is bound up in our current understanding of the migrational history of the solar system. The NICE model \cite{Gomes:2005} has predicted that the Kuiper Belt was disrupted some 600-800 Myrs after the CAI and iron meteorite formation, when Jupiter and Saturn moved into a 2:1 resonance which forced an inward migration while Uranus and Neptune moved outward. This planetary migration greatly disrupted the orbits of the Kuiper Belt planetesimals, scattering some 99\% of them inward or outward in the solar system and creating the system-wide Late Heavy Bombardment, while sweeping up a small fraction ($\sim$100 are known today) into mean motion resonances with Neptune (e.g., the Plutinos). It is thought that while this was happening, some of the KBOs scattered inward would be captured into the stable Lagrangian points around the giant planets. Thus comparing the makeup of the Jovian Trojan population to what is known about KBOs (and other purportedly captured KBOs, like Saturn's moon Phoebe and Neptune's moon Triton) is an important test of this model for solar system development. Other possibilities for the Trojans' sourcing are capture of outer main belt asteroids, or formation of the Trojans in situ from the PPD as Jupiter formed. 

SPHEREx will be able to test each of these hypotheses by comparing the Trojan spectral results to its MBA spectra catalogue. SPHEREx would also be able to compare these Jupiter Trojan spectra to spectra of more distant asteroidal bodies linked with the KBOs, like the Centaur and SDO populations. SPHEREx is likely to measure several of these bodies.

The nature of the Trojan and Greek asteroids is thought to be such an important issue that in 2016 NASA selected the LUCY 5-Trojan tour mission as one of its next DISCOVERY program missions. The 5 objects were selected for their diversity in size, albedo, and location amongst the Trojan population, to hopefully provide a good sampling of the different kinds of source (i.e. captured, formed in place, cometary, scattered MBA) population. The mission’s science goals require Earth-based surveys to put these targets in detailed context of the larger overall Trojan population, and SPHEREx is excellently qualified to do this. The project has contacted the LUCY PI and will be working with the team to help deliver the required context measurements from the Legacy catalog, and to help the team select new extended mission flyby targets.

\section{TESS and Gaia}

\subsection{Improved Knowledge of Exoplanet Host Stars and Exoplanet Properties}

The transit and radial velocity (RV) exoplanet detection techniques have led to the discovery of thousands of exoplanets (approximately 3,500 according to the NASA Exoplanet Archive as of 26 Feb 2018), and upcoming surveys such as the Transiting Exoplanets Survey Satellite (TESS) mission (successfully launched in April 2018) are expected to find thousands more \cite{Sullivan:2015}. The masses and radii measured for exoplanets discovered with these methods depend directly on those of their host stars.

Observations of transiting planets provide our best information on the properties of exoplanets. By combining SPHEREx spectra and Gaia observations, both spectrophotometric and astrometric, we will obtain the radii of the TESS target stars with unprecedented precision. The radii of transiting exoplanets are determined by their measured transit depths (which are proportional to $[R_{planet}/R_{star}]^2$), for which the uncertainties in $R_{star}$ may predominate. The SPHEREx/Gaia data, by constraining $R_{star}$, will translate to measurements of transiting planet radii with 1\% precision, which will be necessary to exploit fully the wealth of data returned by TESS, PLATO, ARIEL, and future exoplanet studies. The improved radii enabled by SPHEREx will be particularly useful for planets orbiting cool K and M stars – which are often targeted in transiting exoplanet studies because of their small sizes, deeper transits and smaller habitable zones. As a pair of all-sky surveys, SPHEREx and Gaia will provide complementary data about all host stars for transiting planets including the targets of future missions, most notably ESA’s PLATO, and also for ground-based transit surveys, such as SPECULOOS, the successor to TRAPPIST. Indeed, Gaia will provide distances to many interesting stars of all types identified by SPHEREx, including 100s of the late M dwarfs and L-type brown dwarfs discussed below. 

Determining stellar parameters to high precision will facilitate detailed studies of exoplanet properties.  For example, knowledge of a rocky planet's measured bulk density alone is insufficient to uniquely constrain its interior composition, so planetary interior models often use stellar elemental abundances, such as the silicon-to-iron ratio, to further constrain the planet's composition.  Empirical measurements of the star-planet [Si/Fe] correlation can improve the accuracy of these models, but such measurements require percent-level planetary radius and mass measurements \cite{Wolfgang:2018}. These requirements necessitate percent-level precision for the stellar parameters. Low-mass stars such as M dwarfs provide the best opportunities for finding small, rocky planets; for a given planet, a smaller, less massive star yields larger radius and mass ratios and thus larger transit and RV signals. However, current models are discrepant at the few-to-several percent level with observations.

\citet{Stevens:2017} and \citet{Beatty:2017} argue that precise and accurate stellar and planetary parameters will soon be measurable by a joint analysis of transit photometry, RV measurements, stellar broad-band flux measurements, and ultra-precise Gaia parallaxes, combined with SPHEREx spectrophotometry. For a typical TESS target with 8 $< V <$ 11, a wealth of broad-band photometry exists in the literature from the UV (GALEX) to the IR (WISE). However, for dwarfs with spectral types M through A, only 40-70\% of the flux is captured in these measurements. This type of analysis therefore requires the use of model atmospheres to infer the stellar bolometric flux from the spectral energy distribution (SED); the bolometric flux, combined with a parallax, provides the stellar radius, and the other stellar and planetary parameters can then be determined from the radius and the transit and RV data. In this case, the precision and accuracy of the inferred stellar and planetary properties hinge on the precision and accuracy of stellar atmosphere models.

The latest Gaia release (DR2) provides, in addition to parallaxes, low-resolution spectrophotometry from 330-1050 nm, covering the stellar spectral energy distribution (SED) peak for the earlier spectral types.  SPHEREx's VIS-near-IR spectrophotometry between 0.75-5.0 $\mu$m would cover the SED peak for mid-K and later spectral types, as well as the Rayleigh-Jeans tail of all observed stars. Adding these measurements to the existing flux measurements then captures 90-98\% of the total stellar flux for these spectral types, enabling bolometric flux determinations that are either independent or only weakly dependent on model atmospheres.

\begin{table}[tbp]
\centering
\begin{tabular}{|c|c|c|c|c|}
\hline
\hline
Spectral  & & \multicolumn{3}{c|}{Values, Uncertainties, \& Fractional Precision}\\
\cline{3-5}
Type & Parameter & Broadband & Broadband \&  & Broadband, Gaia, 
\\
\& Magnitude & & Photometry Only & Gaia Photometry & \& SPHEREx Photometry\\
\hline
& $T_{eff}$ (K) & 	6300 $\pm$ 100 (2\%) & 6290 $\pm$ 50 (0.8\%) & 6265 $\pm$ 15 (0.2\%) \\
\cline{2-5}
F Dwarf  & $F_{bol}$ & 3$\times 10^{-9} $ &	$3\times 10^{-9} $  &	$3\times 10^{-9}$ \\
& (erg/s/cm$^2$) & $ \pm 10^{-10}$ (3\%) & $\pm 5 \times 10^{-11}$ (2\%) & $\pm 2\times 10^{-11}$ (0.6\%) \\
\cline{2-5}
(V $\simeq$ 10) & $R_\star (R_{\odot})$ & 1.59 $\pm$ 0.04 (2\%) & 1.62 $\pm$ 0.02 (1\%) & 1.589 $\pm$ 0.005 (0.3\%) \\
\hline
& $T_{eff}$ (K)	& 3900 $\pm$ 50 (1\%) &	3888 $\pm$ 25 (0.6\%) &	3912 $\pm$ 8 (0.2\%)\\
\cline{2-5}
M Dwarf & $F_{bol}$ & 5$\times 10^{-11}$  &	5$\times 10^{-11}$  &	$4\times 10^{-11} $ \\
& (erg/s/cm$^2$) & $\pm 2\times 10^{-12}$ (4\%)& $\pm 9\times 10^{-13}$ (2\%)& $\pm 3 \times 10^{-13}$ (0.8\%)\\
\cline{2-5}
(V $\simeq$ 16) & $R_\star (R_\odot)$ & 0.58 $\pm$ 0.03 (5\%) & 0.582 $\pm$ 0.014 (2\%) & 0.570 $\pm$ 0.005 (0.9\%) \\
\hline
\hline
\end{tabular}
\caption{Precision on the stellar effective temperature $T_{eff}$, bolometric flux $F_{bol}$, and radius R$_\star$* inferred from SED modeling of two  planet-host stars: an F dwarf (KELT-3; \citet{Pepper:2013}), and an M dwarf (NGTS-1; \citet{Bayliss:2018}). From left to right, the columns show results for an SED fit to extant broadband photometry only, broadband photometry combined with Gaia spectrophotometry, and broadband photometry combined with both Gaia and SPHEREx spectrophotometry.  In each case, a precise distance determination from Gaia is assumed.}
\label{tab:tess_r}
\end{table}

\begin{figure}[!th]
\centering
\includegraphics[width=0.8\textwidth,angle=0]{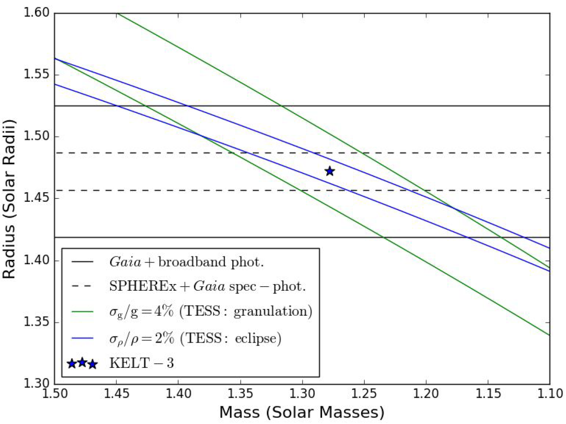}
\caption{Estimated mass and radius constraints on the transiting hot Jupiter host KELT-3. The star denotes the mass and radius values from \citet{Pepper:2013}. The constraint from a 4\% surface gravity inferred from granulation-driven flux variations observable with TESS' short-cadence photometry (e.g. \citet{Kallinger:2016} is shown in green, while a 2\% stellar density constraint from TESS transit observations at the short cadence is shown in blue. The radius constraint from SED modeling using only broad-band flux measurements and a Gaia parallax with 7 micro-arcsecond uncertainty is shown by the black solid lines, while the more precise constraint achieved by incorporating SPHEREx and Gaia spectrophotometry is shown by the dashed black lines. SPHEREx deep fields overlap with LSST and Euclid deep fields.}
\label{fig:tess}
\end{figure}

Table~\ref{tab:tess_r} shows how, for two representative stars, combining SPHEREx and Gaia spectrophotometry can improve the precision of the SED-derived stellar parameters over fitting to the broad-band photometry alone. By measuring nearly all the stellar flux, effective temperatures can be inferred to $<$10 K ($<1$\%) precision.  Likewise, stellar radii can be determined to $<$ 1\% precision, a level comparable to that achieved with interferometric techniques, but this method is applicable to stars hundreds of parsecs away.  Figure~\ref{fig:tess} illustrates the power of this stellar radius constraint – both in terms of the radius precision itself and how this constraint complements the high-precision density constraints achievable with short-cadence observations from TESS.  Note that the broad wavelength coverage of the Gaia-SPHEREx data set also allows determination of and correction for interstellar reddening even for extinction as low as $A_v \simeq$ 0.05, which further improves the precision of the determination of stellar parameters.  With the stellar parameters so well-determined, those of the transiting planet will be limited only by the uncertainties in the measurements of transit depth and stellar RV.

TESS is expected to provide high-quality transit data for $\simeq$ 1,700 planets, of which $\simeq$ 550 are expected to be Earths and Super-Earths and to detection antoerh $\sim$20,000 giant exoplanets in the full frame image data \cite{Sullivan:2015}. Gaia and SPHEREx will produce an all-sky archive which will be available to characterize not only all TESS exoplanet host stars but also almost all identified in the past and the future from both ground and space observatories.  For example, ESA’s PLAnetary Transits and Occultations (PLATO) satellite – to be launched in 2026 - is expected to characterize ~4,800 V $<$ 13 transiting systems precisely, targeting longer-period systems to search for Earth-like planets in the habitable zones of Sun-like stars, of which it is expected to find $\simeq$ 60 (ESA 2017).  Like TESS, PLATO will concentrate on bright stars readily measured by SPHEREx and Gaia.  With the abundance of high-precision stellar densities from the transit and asteroseismology measurements from these satellites, SPHEREx and Gaia have the potential to improve our knowledge of thousands of stellar and planetary radii and masses, ushering in an area of high precision exoplanet studies.

\subsection{Stellar Astronomy}

As all sky surveys with contiguous spectral coverage, SPHEREx and Gaia will undoubtedly prove synergistic in many, many ways.  Consideration of the forgoing discussion, for example, tells us that the pair will constrain the properties of stars in a way which may ultimately distinguish between different sets of evolutionary tracks.  The improved stellar data will be particularly valuable when paired with the higher resolution spectroscopic data to be obtained by a number of ground-based spectroscopic surveys now getting underway (see \S~\ref{sec:spectro_surveys}).



\section{eROSITA and Clusters of Galaxies}

eROSITA (extended ROentgen Survey with an Imaging Telescope Array) is  the primary instrument on the Russian Spektrum-Roentgen-Gamma (SRG) mission. eROSITA will provide an all-sky Xray survey every 6 months for 4 years, with a  final expected depth of $1\times 10^{-14}$ ($3\times 10^{^-15}$ erg/cm$^2$/s at the ecliptic poles) which is about 30 times deeper than ROSAT \cite{Voges:1999,Boller:2016} in the soft band (0.5-2 keV). eROSITA will also provide for the  first time ever an all-sky survey in the hard band (2-10 keV), reaching an expected depth of $2\times 10^{-13}$ erg/cm$^{2}$/s  ($4\times 10^{-14}$ erg/cm$^2$/s at the ecliptic poles). The scientific main driver of eROSITA is cluster cosmology, and in particular, the growth of the Large Scale Structure. It is expected to detect $10^5$ massive clusters of galaxies up to redshift $\simeq$ 1. eROSITA  will also detected about 3 million AGN to z$\simeq 6$, and about 500,000 stars.

\begin{figure}[!th]
\centering
\includegraphics[width=0.8\textwidth,angle=0]{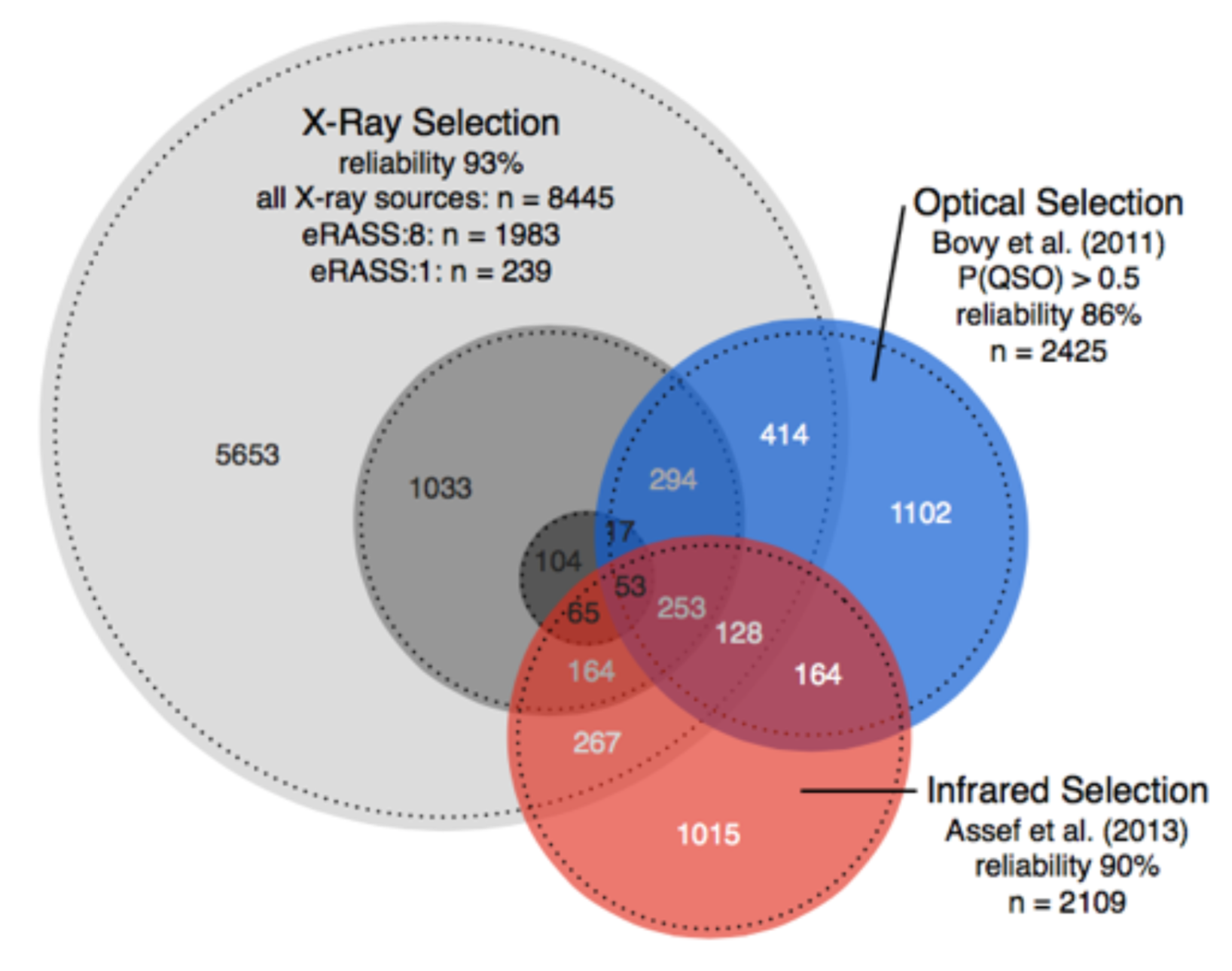}
\caption{This figure (from \cite{Menzel:2016}) illustrates overlap and synergism of infrared and other techniques of identifying AGN, from Mara Salvato presentation at workshop.}
\label{fig:eROSITA1}
\end{figure}

For the complete identification of extended (i.e. clusters) and point-like (i.e. AGN and stars) eROSITA sources, deep multi-wavelength coverage of the sky is essential. In particular SPHEREx will be crucial for the following reasons:
\begin{itemize}
\item SPHEREx will cover the entire sky, and provide full-sky object catalogs with higher angular resolution than eROSITA and homogeneous coverage. Homogeneity on wide areas is fundamental when selecting the counterpart to both point-like and extended X-ray sources. This is particularly true for point-like sources for which the counterpart is defined  in a probabilistic way, with the probability also depending on the number density of the sources (e.g. NWAY, \citet{Salvato:2018, Dwelly:2017}). 
\item SPHEREx will cover the longer optical wavelengths, NIR and MIR. At these wavelength most AGNs are easily identified, either due to their intrinsic spectral energy distribution (SED) \cite{Yamamura:2010}, or due to the redshifting of the strong UV emission of high-redshift AGNs into the SPHEREx bands. At the same time the number of field sources emitting at these wavelength  decreases, thus improving the success rate of counterpart identification methods.
\end{itemize}

\begin{figure}[!th]
\centering
\includegraphics[width=0.7\textwidth,angle=0]{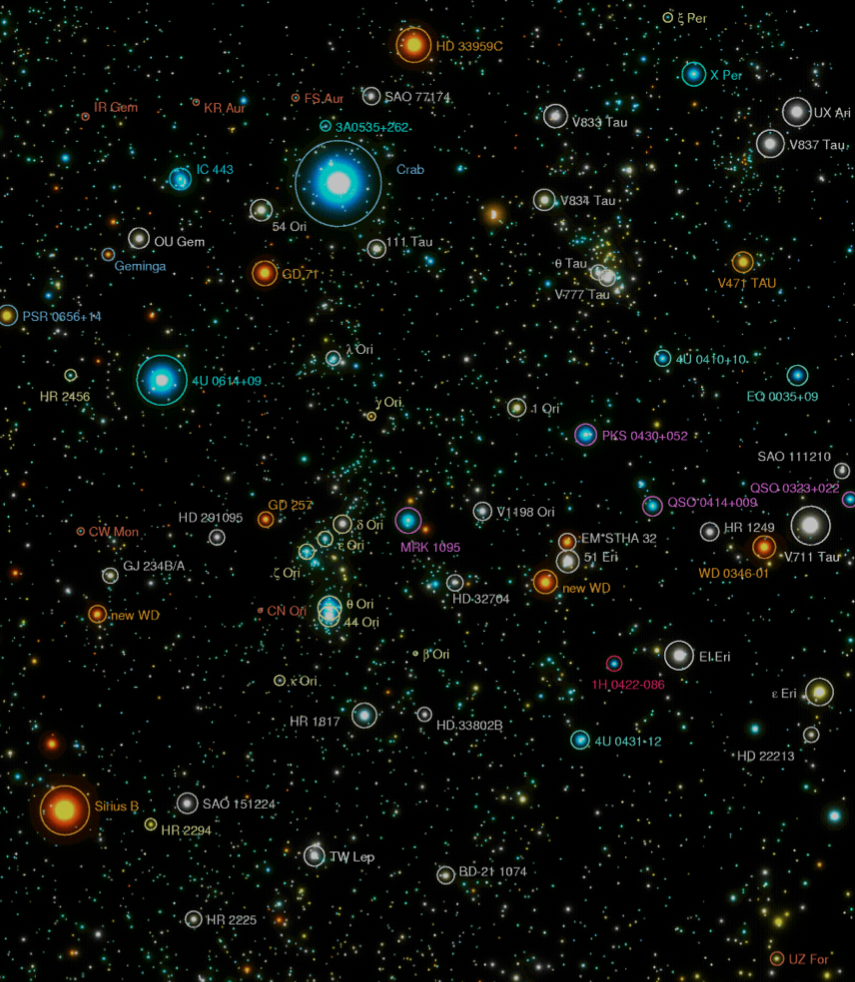}
\caption{This composite image (Courtesy:K. Dennerl), derived from data obtained during the ROSAT all-sky survey, color-coded by their X-ray colors, illustrates the rich variety of sources which can be detected in X-rays: stars, white dwarfs, neutron stars, supernova remnants, active galactic
nuclei, galaxy clusters, and even solar system objects like our Moon. eROSITA, combined with SPHEREx will be sensitive not only for nearby stars such as shown here, but for obscured stars lying in distant regions of the Milky Way.}
\label{fig:eROSITA2}
\end{figure}

Based on magnitude distributions of the counterparts to the X-ray point-like sources detected in STRIPE82x \cite{Lamassa:2016,Ananna:2017}, we expected that at the depth of eROSITA, SPHEREx will provide a reliable counterpart to  90\% and 87\% of the sources  in the integrated K and [3.6 $\mu$m bands, respectively. At the ecliptic poles where both eROSITA and SPHEREx are deeper, virtually all the point-like eROSITA sources will have a counterpart.  Similarly, the preponderance of galaxy members of clusters up to $z \sim 1$ are expected to be detected in the narrow bands of SPHEREx redder than the K band based (see \textit{e.g.} \citet{Lin:2006}).

For the galaxies in the clusters and AGN SPHEREx will be also able to provide an accurate estimate of the photometric redshift by pin pointing the emission/absorption lines  and the  key features in the continuum. The impact that the availability of intermediate and narrow band photometry has on the precision of photometric redshift has been already demonstrated in previous surveys as COMBO-17 \cite{Wolf:2004}, COSMOS-21 \cite{Taniguchi:2005}, SHARDS \cite{Perez-Gonzalez:2013, Cava:2015} and further discussed in \citet{Salvato:2018}.  When the sources will be faint, the photometry will need to be complemented with other deep, ancillary,  optical broad band photometry (see \S~\ref{Sec:photoz}). In particular for AGN, the emission lines from the host and from the vicinity of the central BH, combined with proper SEDs, will allow the determination of accurate photometric redshifts \cite{Ananna:2017, Hsu:2014, Fotopoulou:2012, Cardamone:2010, Salvato:2011}. It will bring much improvement on, for example, the X-ray AGN Luminosity Function, which is still very uncertain at z $>$ 3 e.g., \cite{Marchesi:2016, Civano:2011, Georgakakis:2015}.

SPHEREx will continuously revisit the sky with a six month cadence (at a given position and given wavelength element; and 30 times more frequently at the ecliptic poles). This multi-epoch data set will yield a new understanding of the physics regulating the activity of black holes, and will provide valuable additional information on X-ray sources with ambiguous optical associations through time variability studies.



\subsection{Clusters of Galaxies}

\begin{figure}[!t]
\centering
\includegraphics[width=0.58\textwidth,angle=0]{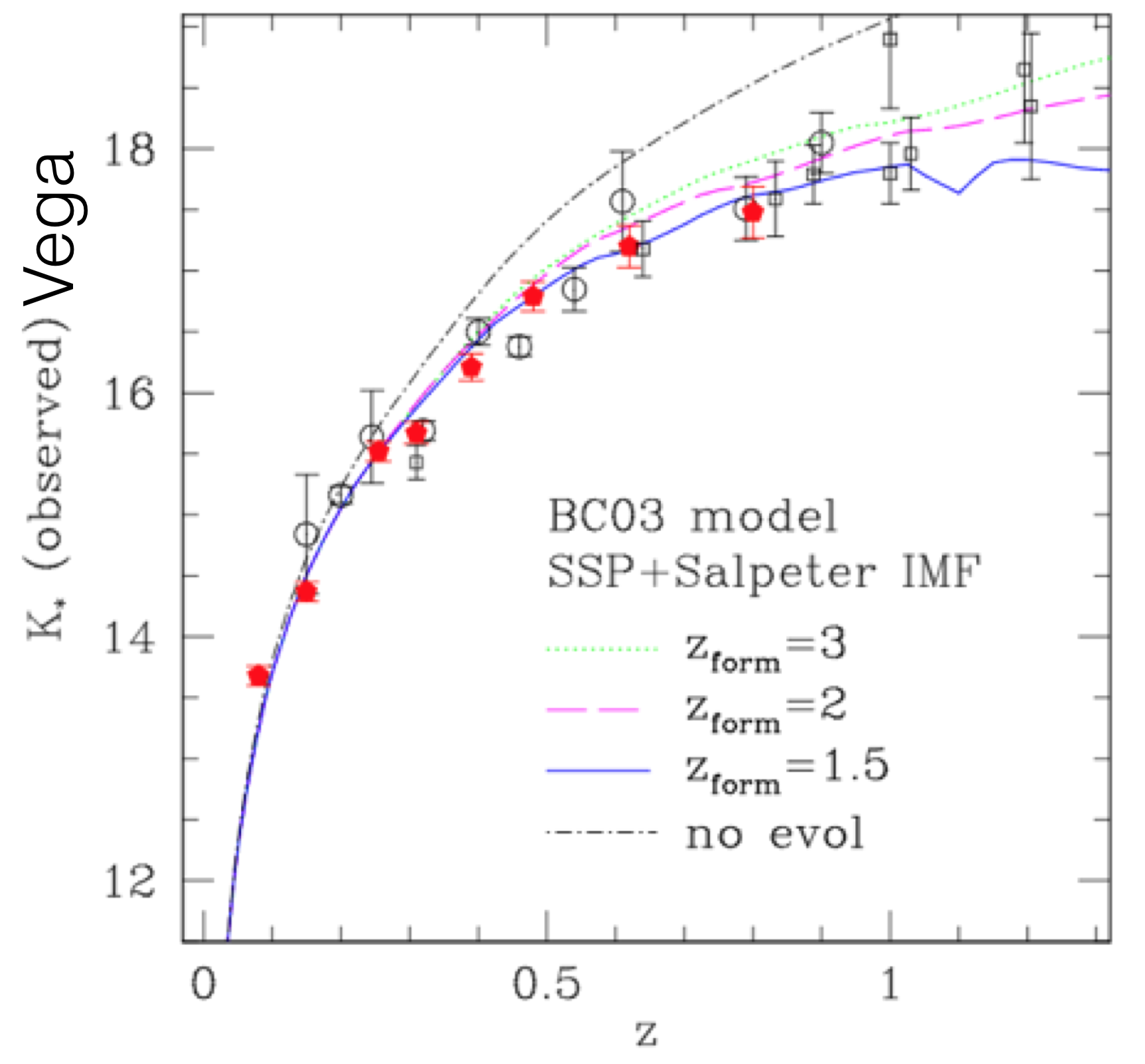}
\caption{This figure (from \citet{Lin:2006}) shows the expected K band magnitude distribution of galaxies  members of clusters up to redshift $\sim$ 1.  SPHEREx can detect individual galaxies in eROSITA discovered clusters out to $z\simeq$ 0.5. The SPHEREx sensitivity at $K_\star$ is $\sim$ 13.5 Vega magnitude.}
\label{fig:clusters_1}
\end{figure}

eROSITA also shows how SPHEREx data complement X-ray and mm-wave observations of clusters of galaxies by providing cluster redshift and stellar mass measurements. Upcoming millimeter-wave cluster surveys (e.g., SPT-3G, AdvACTPol, Simons Observatory, CMB-S4) and the eROSITA X-ray mission will discover 100,000 massive systems \cite{Merloni:2012,Benson:2014,Henderson:2016}. Since these surveys contain limited-to-no redshift information, auxiliary data are required to obtain robust redshifts. Simulations suggest that the precision of cluster redshift measurement from SPHEREx, over the full sky, should equal or exceed that of current generation optical surveys for redshifts z $<$ 0.6 and remain to $\sigma(z)/(1+z) <$ 0.03 for z $<$ 0.9. Furthermore, SPHEREx measurements of the stellar content of galaxies will provide virial masses for large statistical cluster samples. Finally, extrapolating from the SDSS sample \cite{Rykoff:2013}, we forecast that the SPHEREx all-sky maps will independently reveal $\simeq$ 30,000 galaxy clusters.

\section{ALMA}

\subsection{Circumstellar Dust around Main Sequence Stars.}

IRAS made the first detections of what we now call “planetary debris disks”, clouds of dust around main sequence stars which are fed by the collisions and decay of exo-asteroids and exo-comets. With the prevalence of exoplanets we now understand that debris disks reflect the properties of the exoplanetary systems they occupy.  Ground-based observations from the NASA Infrared Telescope Facility have recently detected warm dust radiating shortward of 5 $\mu$m around a handful of stars selected because they were known to have cooler circumstellar material. SPHEREx will search over the entire sky for near-infrared excesses due to warm circumstellar dust radiating shortward of 5 $\mu$m.  These studies are synergistic with both JWST and the ALMA submillimeter array which, respectively, can assay and in many cases image the amount of cooler and cold dust orbiting these same stars. This will provide not only a complete picture of the distribution of circumstellar dust but also glimpses of the underlying planetary systems.  

\begin{figure}[!th]
\centering
\includegraphics[width=0.85\textwidth,angle=0]{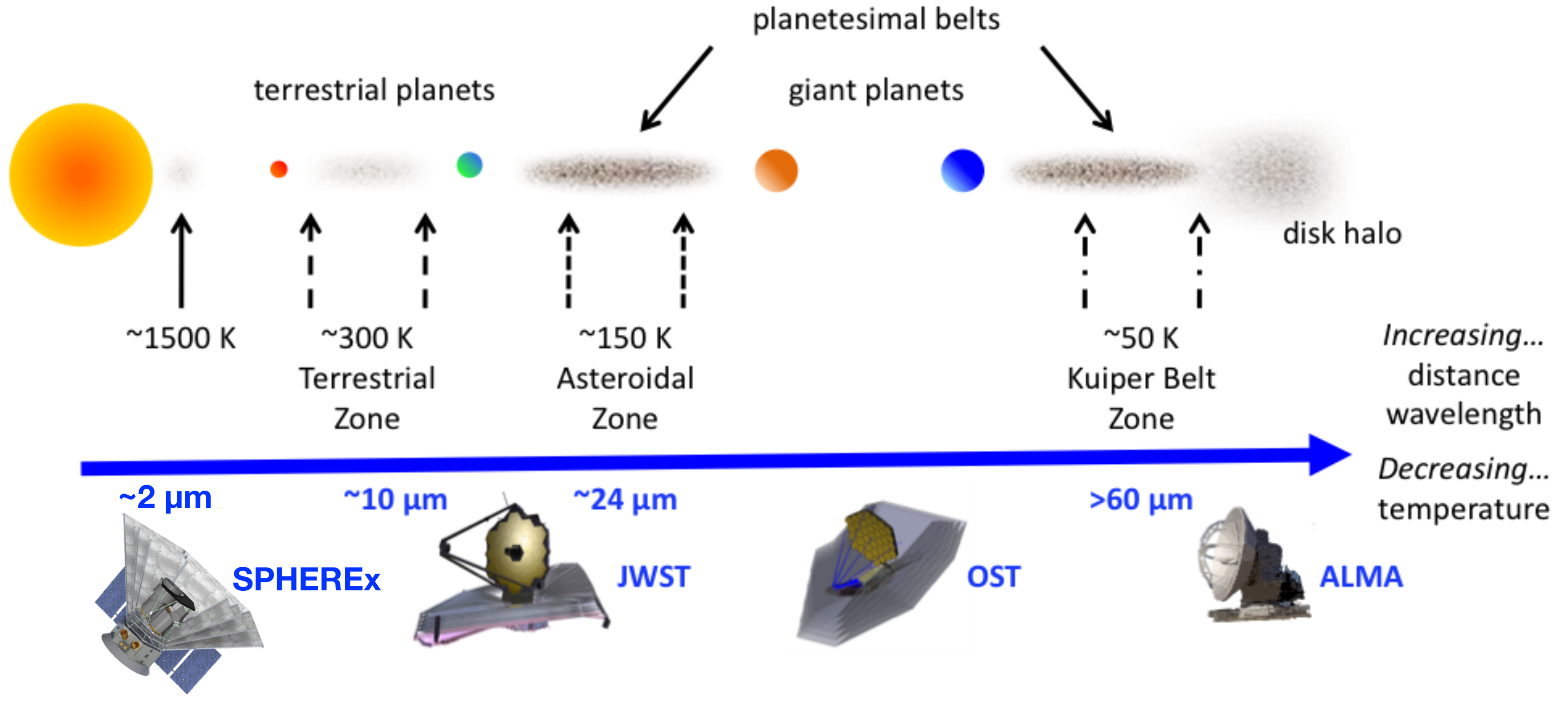}
\caption{SPHEREx complements ALMA and JWST}
\label{fig:alma_v1}
\end{figure}

\subsection{Protoplanetary Disks and Young Stellar Objects}

SPHEREx, JWST, and ALMA will also work together in the study of protoplanetary disks, which are the birthplaces of exo-planets and exo-planetary systems. SPHEREx will be a valuable probe of the innermost, terrestrial planet-forming regions of protoplanetary disks, providing needed synergy with current and future data from millimeter observatories, such as ALMA, that provide information on the distribution of gas and large dust grains down to about 3~AU in nearby ($\sim$140 pc) disks.
 


\begin{figure}[!th]
\centering
\includegraphics[width=0.7\textwidth,angle=0]{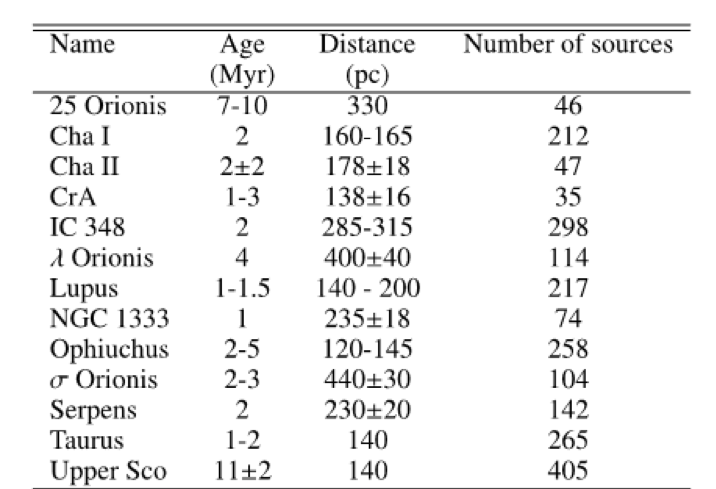} \caption{Table adapted from \cite{Ribas:2014} showing young groups and associations lying within 500 pc.}
\label{fig:yso}
\end{figure}

As the table shown in Fig.~\ref{fig:yso}, adapted from \cite{Ribas:2014} shows, young groups and associations lying within 500 pc of the Sun contain hundreds of young stellar objects with ages less than a few million years, and should thus host planet-forming disks.  These proto-planetary disks, like the debris disks discussed above, will contain dust which emits over a wide range of wavelength from the near-infrared to the submillimeter and beyond.

\begin{figure}[!th]
\centering
\includegraphics[width=0.8\textwidth,angle=0]{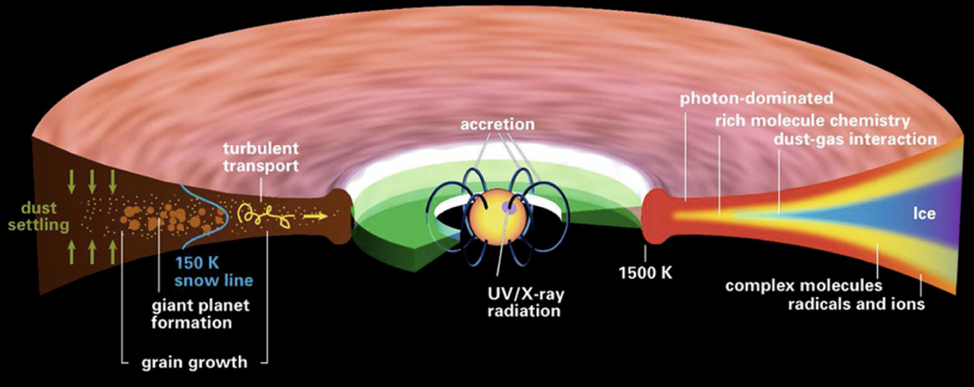}
\caption{SPHEREx complements ALMA and JWST}
\label{fig:yso2}
\end{figure}

The short wavelengths and warm dust sampled by SPHEREx lie in the innermost regions of the disk, just outside the dust sublimation radius, where observations and theory suggest that an inner rim or wall puffs up due to heating by the adjacent star.  Although SPHEREx cannot resolve the structure, it is ideally designed for studying the rim, as the SPHEREx wavelength coverage extends from below 1 $\mu$m, where the starlight will dominate the SED, to the longer wavelengths around 2-to-3 $\mu$m, where the radiation of the rim becomes dominant (Fig.~\ref{fig:yso2}).

This inner rim is of particular importance because it is the surface at which the energy from the star comes into the inner disk, the region from which material on its way to being accreted on to the star begins its inward spiral, and the site of the formation of the inner most planets, perhaps similar to Mercury in the Solar System.  SPHEREx, by elucidating the energetics and geometry of hundreds of thousands of such structures, can greatly increase our understanding of the importance of these and similar processes which occur in the inner regions of forming solar systems.  As was the case for the planetary debris disks, the SPHEREx observations will also identify particularly interesting objects for investigation by JWST and ALMA, both of which can study not only the dust but also the gas in proto-planetary systems.
 
SPHEREx will also provide useful information on the variability of these objects.  About 60--70$\%$ of disk systems are variable in the NIR \citep{morales11, flaherty13}.  The variability observed points to changes in the height of the inner disk rim which may be due to warps and perturbations in the inner disk stemming from interactions of the disk with a planet or the stellar magnetic field \citep{muzerolle09,espaillat11,flaherty11,stauffer14}. 

\begin{figure}[!b]
\centering
\includegraphics[width=0.48\textwidth,height=0.3\textwidth,angle=0]{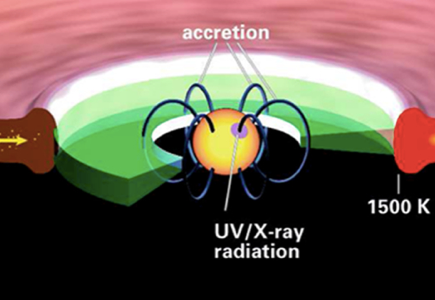}
\includegraphics[width=0.48\textwidth,height=0.3\textwidth,angle=0]{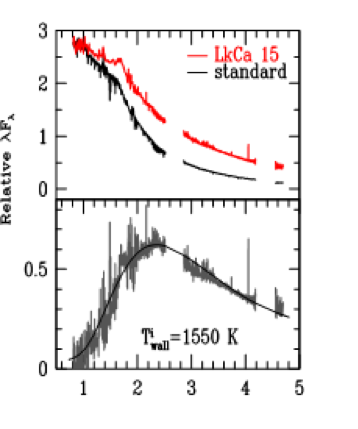}
\caption{The excess near infrared emission seen by SPHEREx - shown in red in the panel at the upper right and then with the standard model subtracted in the lower right can be used to study the geometry of the puffed up inner wall of an accretion disk around a Young Stellar Object.}
\label{fig:yso3}
\end{figure}

\section{SOFIA}

With its database of sources with medium resolution IR spectra over the entire sky, SPHEREx will provide a rich source of targets for follow-on studies at wavelengths longer than 5 microns. For many of these targets, follow-on studies in the wavelength range from 5 to 320 microns are essential for gleaning key quantitative information on the nature and evolutionary state of these sources: from high redshift galaxies to local planet-forming disks around young stars and protostars. 

\begin{figure}[!th]
\centering
\includegraphics[width=0.98\textwidth,angle=0]{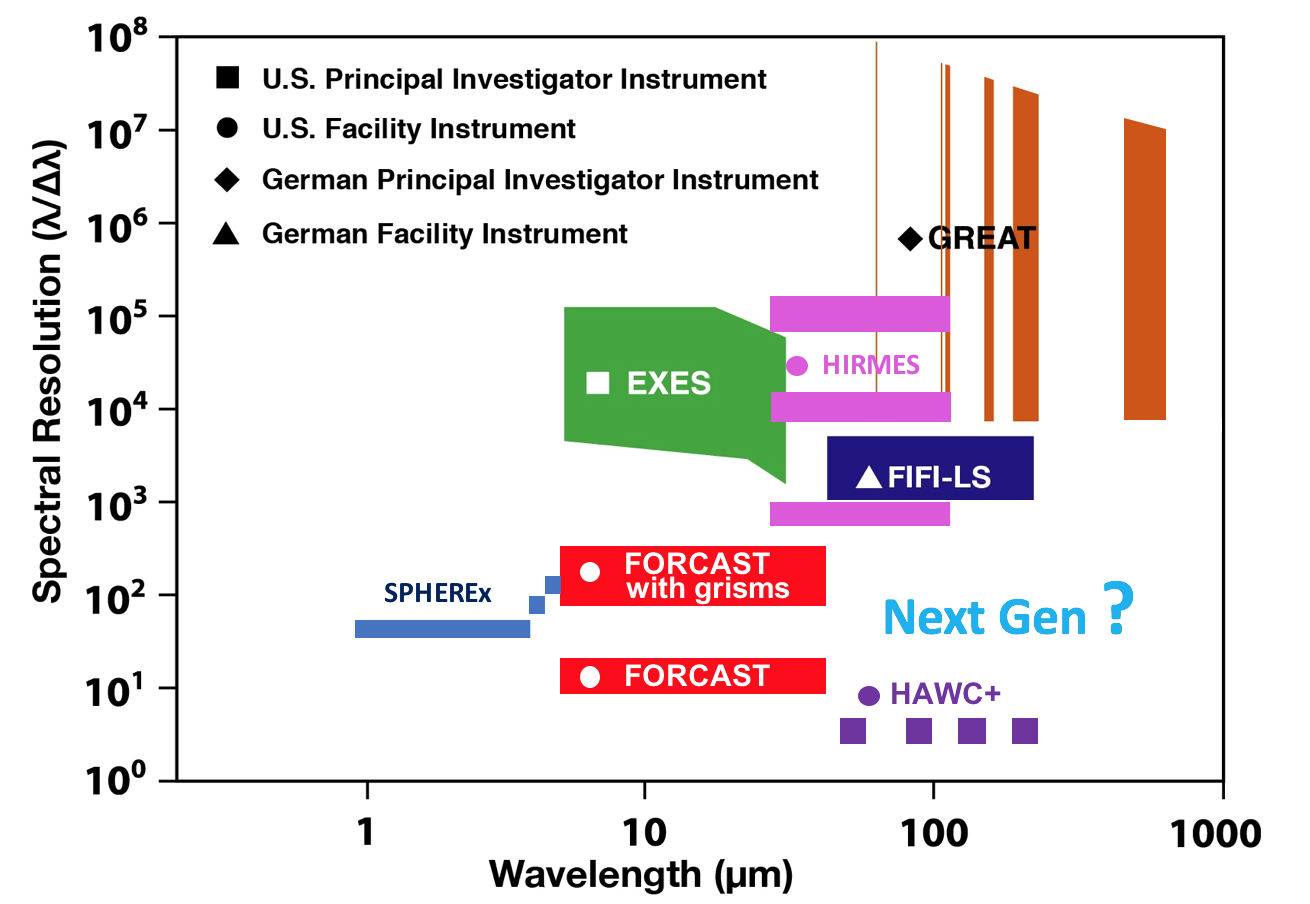}
\caption{SOFIA’s near future instrument suite on the spectral resolution versus wavelength plane compared to SPHEREx. It is likely that several of SOFIA’s instruments will be replaced with different but more capable instruments by the time of SPHEREx’s launch.}
\label{fig:sofia}
\end{figure}

In particular, SPHEREx will provide a complete all-sky collection of distant galaxies brightened by gravitational lensing. SOFIA’s unique wavelength coverage (Fig.~\ref{fig:sofia}) will allow the determination of parts of these galaxies’ SEDs otherwise inaccessible for at least a decade. Recently with SOFIA-HAWC+, Riechers, et al. 2018 (in prep) observed the mid-infrared SED of a z=3.9 galaxy (rest wavelengths 11, 18, \& 31 microns) and Ma et al., (2018, submitted) observed a z=1.03 galaxy at rest wavelengths 26, 44, \& 76 microns. These high S/N measurements at crucial wavelengths would allow the determination of AGN versus starburst activity as a function of redshift and bolometric luminosity with a sufficiently large sample size. The current sample size of gravitational lens enhanced galaxies is of the order of a few tens, whereas SPHEREx should supply a sample of over hundreds of SOFIA-accessible targets. For the two cases cited above, all three wavelengths were observed with a S/N exceeding 10 in less than one hour.

A second case of interesting follow-on studies are time-dependent phenomena discovered by SPHEREx. It is now generally recognized that protostars of all masses accrete material from their surrounding circumstellar disks stochastically rather than growing at a steady rate. Little is known about such accretion bursts, but it is clear that accretion bursts will result in a variety of related phenomena: luminosity outbursts, variable feedback (radiation and mechanical), variability in outflows and jets. Because of variable accretion and sudden accretion bursts, the outer layers of the central protostars are likely not in thermal equilibrium (one of the basic assumptions of stellar structure models). As a single example, \cite{Caratti:2017} used FORCAST and FIFI-LS observations to derive fundamental parameters of an accretion burst in the high-mass YSO S255IR-NIRS 3. Without these far infrared measurements, quantitative studies of this phenomenon would not have been possible. SPHEREx will provide SOFIA with hundreds of potential targets for related studies, which allow us to finally understand this basic mechanism of star formation.

SPHEREx will find all regions containing ices that will require follow-on studies by SOFIA's HIRMES instrument, specifically designed and optimized to study water in both gaseous and in its solid-state in different phases of planet formation. HIRMES will also be able to observe emission from the lowest excited states of both H2 at 28 microns and HD at 112 microns. Supplementing HIRMES' [O I] 63 micron measurements and [C II] 158 micron studies with FIFI-LS and/or upGREAT can provide crucial information of the structure, temperature and kinematics in protoplanetary disks in different phases of evolution, including high accretion phases discussed above.

SOFIA provides the astronomical community only access to the far infrared wavelength regime for the near future up to its expected lifetime until 2034. Thus, SPHEREx’s spectral surveys and target lists will be a valuable resource to enhance the discovery space of SOFIA during the era of JWST and ALMA.

\section{Large-Scale Ground-based Integral Field Spectroscopy Surveys}
\label{sec:spectro_surveys}

\subsection{A New Generation of Integral Field Spectroscopy Surveys}

In the past few years, a number of major ground-based spectroscopic surveys with great synergism with SPHEREx have been proposed or are currently underway. They cover visible and near infrared wavelengths and are remarkably synergistic with SPHEREx.  A recent publication \cite{Kollmeier:2017} lists close to a dozen such programs expected to be operating prior to and during the SPHEREx mission (Fig.~\ref{fig:mission_wavelength}).  SPHEREx has lower spectral resolution than these surveys, which for the most part aim at studying stellar atmospheres or emission lines in the Milky Way and nearby galaxies.  However, what SPHEREx does uniquely is to extend the wavelength coverage of these surveys into the 2 to 5 $\mu$m region while covering the entire sky so that a SPHEREx spectrum will be available for each of the targets of these other surveys.  As a specific example, we compare SPHEREx with SDSS-V, a ground-based survey program to be carried out in both hemispheres with three distinct components envisioned to be operating in the 2020-2024 time frame.  These three elements are summarized in the following table, adapted from Kollmeier et al. \cite{Kollmeier:2017}, to which we have added a fourth element that they describe in their paper, SPHEREx, and a separate survey called MOONS which is intended to be carried out on the VLT \cite{Cirasuolo:2014}. MOONS is included because it resembles SPHEREx in wavelength coverage.  These surveys rely on large fiber bundles to feed multi-object spectrographs.  When the targets are stars and galaxies, the fibers are positioned appropriately.  For the LVM part of the SDSS survey, they are closely packed to achieve complete areal coverage on the sky.

It is clear from Table~\ref{tab:spectro_survey} that there is a tremendous amount of synergism between SPHEREx and all of these surveys, none of which extends beyond 2 $\mu$m in wavelength. SPHEREx will provide a long wavelength extension to essentially every spectrum obtained by these major programs, as well as a short wavelength overlap which will assure  proper matching of the spectra and cross-calibration.  As is the case with the 2MASS and WISE surveys described earlier, this added spectral coverage will enhance the value of the survey data in ways beyond a simple sum of the surveys. In some cases, particularly that of MOONS, one can also imagine these survey instruments, with their higher spectral resolution, being used to follow up on results from SPHEREx.   More specifically, a few examples of how SPHEREx will connect with the elements of the SDSS-V survey include:
\begin{enumerate}
\item {\bf Milky Way Mapper [MWM].}  A major objective of the MWM is to study reddened stars in the plane of the galaxy.  The addition of SPHEREx data will allow the reddening of these stars to be determined precisely and the stars to be positioned along the line of sight, with reference to the three-dimensional distribution of interstellar dust developed by Finkebeiner and colleagues \cite{Schlafly:2014}, even for stars too highly reddened to be seen by Gaia.
\item {\bf Black Hole Mapper [BHM].}  A major objective of the BHM is to study the structure of AGN by reverberation mapping, which measures the time delay between optical variability of the AGN and the response of the circum-nuclear gas and dust.  SPHEREx will provide constraints on the interpretation of these data by obtaining spectra of these objects which span the complete wavelength range from the variable heating in the optical to the response of the heated dust in the infrared.
\item {\bf Local Volume Mapper [LVM].}  LVM will obtain complete spectra from 0.36-to-1 $\mu$m of selected regions within the Milky Way, the Magellanic Clouds, and Local Group Galaxies.  It will provide data on the gas in these regions to complement, at least locally, the stellar spectra obtained by the MWM.  SPHEREx will add to the LVM data cubes information obtainable only in the infrared:  absorption due to interstellar/circumstellar ices; PAH emission at 3.3 $\mu$m; and gas phase emission lines, not only recombination lines of HI but also emission from shock-excited CO and H2.  The LVM pixels sizes will range between 2.7" and 37", depending on the instrument configuration. Given that it will be easy to degrade the resolution of either survey, these numbers match nicely with the 6" SPHEREx pixels.
\item {\bf The Solar Vicinity Census (SVC).} SVC is a subproject of the MWM aimed at getting the best possible spectra of all stars within 100 pc using both the infrared and the visible wavelength spectrographs which make up the SDSS-V instrumentation.  This census will be greatly augmented by the addition of the [SPHEREx + Gaia] data discussed earlier in the context of exoplanet host stars.  This will provide data on stellar masses, radii, and temperatures  which will combine with the high resolution SVC spectroscopy to provide a complete data base on the nearest stars, which will have enough entries (400,000) to support statistical studies as well as analysis of individual stars.
\end{enumerate}

\subsection{Spectrally Mapping Nearby Galaxies}

A particularly strong synergy between SPHEREx and these coming surveys will be the spectral mapping of nearby galaxies. While SPHEREx was not necessarily designed with nearby galaxy observations in mind, the spatial and spectral coverage will allow many interesting studies.  In particular, in nearby galaxies we can image the individual stars that are producing the light seen in the SPHEREx spectroscopy.

Since the SPHEREx spectra covering nearby galaxies will be the integrated spectra of a large number of stars, it will be informative to compare these to the properties of the resolved stellar populations whenever possible.  With the current HST resolved star library on nearby galaxies, some of this kind of work can be done, but when WFIRST gives us full coverage of these galaxies, there will be a wealth of information for testing population synthesis models and mapping galaxy properties.  Deeper exposures than the nominal SPHEREx all sky survey at the locations of large galaxies within 10 Mpc would be of great interest for such work.

Matching integrated spectroscopy and resolved stellar photometry has been ongoing in the optical, and has yielded encouraging results, finding general agreement between models and data in many cases \cite{Johnson:2013,Conroy:2013, Byler:2017}, but there are also offsets that exist.  A large amount of such optical spectroscopy is available through SDSS IV, and more is planned in SDSS V. The SPHEREx spectroscopy will complement all of this effort in the optical with NIR spectra, providing a huge library of data for improving population synthesis models in the NIR.   In particular, the near IR can be difficult to model reliably because of high luminosity and short-lived red He-burning stars and asymptotic giant branch stars \cite{Girardi:2010,Melbourne:2012,Boyer:2013}.  There is still much work to do to understand the effects of these, especially when interpreting NIR spectroscopy, and SPHEREx and WFIRST together should allow large leaps in our understanding.

\begin{table}[tbp]
\centering
\begin{tabular}{|c|c|c|c|}
\hline
\hline
Program & Science Targets &	$N_{objects}$ &	Wavelength Range,  \\
& &	and Sky Area &	Resolution,	\\
& &	and Sky Area &	and Limiting Sensitivity\\
\hline
Milky Way Mapper & Stars across the MW, & $>$6M stars;  & 1.51-1.7 $\mu$m,\\
(MWM, part of SDSS-V) & esp. in the plane & 4 $\pi$ sr & R=22000, \\
& & & H$\simeq$13.4  \\
\hline
Black Hole Mapper & Supermassive Black Holes	& $>$400,000; 4$\pi$ sr. & 0.37-1 $\mu$m\\
(BHM, part of SDSS-V) & & & R=2000 \\
& & & I$\simeq$20  \\
\hline
Local Volume Mapper &	ISM and stars in MW  & 	$>$ 25M spectra  & 0.36-1 $\mu$m \\
(LVM, part of SDSS-V) & and and nearby galaxies & contiguous over $\simeq$ 1 sr. & R=2500-4000 \\
\hline
Solar Vicinity Census &  All stars with 100 pc	 & Combines Instruments & G$\simeq$20, H$\simeq$12 \\
(SVC, part of SDSS-V) & 400,000 stars, 4pi sr. & used in rows 1\&2  &\\
\hline
MOONS  & Varied &	10,000,000 objects, & 0.6-1.8 $\mu$m,\\
ESO initiative on VLT & & $\simeq$ .05 sr. & R=5000-20000,\\
& & & G$\simeq$22, H$\simeq$17\\
\hline
SPHEREx	& Agnostic & $>$ 100,000,000 stars, & 0.75-5 $\mu$m,\\
& &  $>10^9$ galaxies, 4$\pi$ sr. & R=30-130,\\
& &  6" pixels & H$\simeq$19, I$\simeq$18 \\
\hline
\hline
\end{tabular}
\caption{\label{tab:spectro_survey}}
\end{table}

\section{SPHEREx Observation of the Interstellar Medium at High Redshift}

\subsection{ISM masses at High Redshift} 

SPHEREx will also allow us to probe directly the multiple phases of Hydrogen. We first review in very qualitative fashion the ISM contents of star forming (SF) galaxies at high redshift as derived from ALMA measurements the Rayleigh-Jeans dust continuum. The translation of RJ  continuum fluxes at $\lambda_{rest} > 300 \mu$m to ISM masses is based on an empirical calibration derived from a sample of over 75 low-$z$ normal SF galaxies, ULIRGs and high-$z$ sub-mm galaxies (SMGs) having 
measured RJ dust and CO (1-0) fluxes. This calibration shows better than factor 2 accuracy (see \cite{sco17}). The empirical calibration of the technique \citep{sco17} is based on three different low and high redshift galaxy samples: (\textit{i}) a sample of 30 local star forming galaxies; (\textit{ii}) 12 low-$z$ Ultraluminous Infrared Galaxies (ULIRGs);  and (\textit{iii}) 30 $z \sim 2$ SMGs. 

ALMA observations of the long wavelength dust continuum were used to estimate the interstellar medium (ISM) masses in a sample of 708 galaxies  at $z = 0.3$ to 4.5 in the COSMOS field \citep{sco17}. The galaxy sample has known far-infrared luminosities and, hence, star formation rates (SFRs), and stellar masses (M$_{\rm *}$)  from the optical-infrared spectrum fitting. The derived ISM masses are used to determine the dependence of gas mass on redshift, M$_{\rm *}$, and specific SFR (sSFR) relative to the galaxy Main Sequence (MS).  

\begin{figure}[!ht]
\centering
\includegraphics[width=0.7\textwidth]{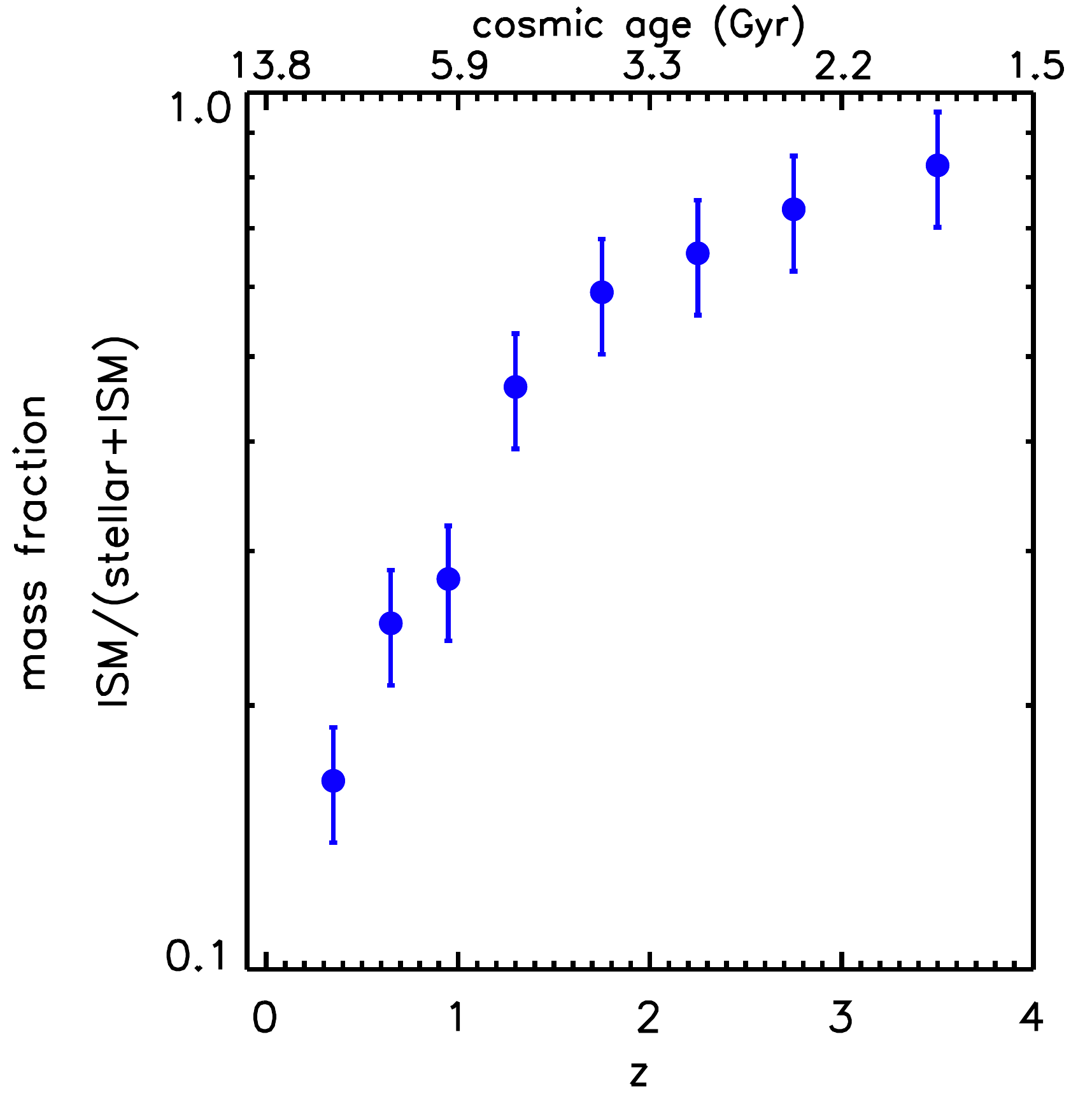}
\caption{The mass fraction of ISM are shown for galaxies with stellar masses $\rm M_{\rm *} = 10^{10}$ to $10^{12}$ M$_\odot$ as derived from the Rayleigh-Jeans dust emission \citep{sco17}}
\label{lilly_madau_2}  
\end{figure}

Enormous increases in the gas contents of galaxies are seen above $z \sim 1$, getting up to gas mass fractions ( gas / (stars + gas)) of $80\%$ at $z = 4$ (see Figure \ref{lilly_madau_2}). The ISM masses increase approximately as the 0.63 power of the rate of increase in SFRs with redshift and the 0.32 power of the sSFR/sSFR$_{MS}$. The SF efficiencies also increase as the 0.36 power of the SFR redshift evolutionary and the 0.7 power of the elevation above the MS; thus the increased activities at early epochs are driven by both increased ISM masses and SF efficiency. Using the derived ISM mass function we estimate the accretion rates of gas required to maintain continuity of the MS evolution ($>100$ \msun yr$^{-1}$ at $z > 2.5$).  Simple power-law dependences are similarly derived for the gas accretion rates.  We argue that the overall evolution of galaxies is driven by the rates of gas accretion. The cosmic evolution of total ISM mass is estimated and linked to the evolution of SF and AGN activity at early epochs.  Table~\ref{equations2} summarizes the ISM contents, gas mass-fractions and accretion rates normalized to $z = 2$.

\begin{table}[tbp]
\centering
\begin{tabular}{l}
\hline
\hline
$\rm M_{\rm ISM}  = \rm 8.65\times 10^{10}~\msun~\times \left[(1+z)_2^{1.84} \times (sSFR/sSFR_{MS})^{0.32} \times  M_{*~5}^{0.30}\right]  $ \\
\\
$\rm SFR = \rm 9.9 \times  \left({M_{\rm ISM}\over{10^{10} \msun}}\right) ~\msun~yr^{-1}~~\times  \left[(1+z)_2^{1.04} \times  (sSFR/sSFR_{MS})^{0.70} \times  M_{*~5} ^{0.01} \right]$ \\
$\rm ~~~~  ~~~ = \rm 85   ~\msun~yr^{-1}~~\times  \left[\left({M_{\rm ISM}\over{8.65\times10^{10} \msun}}\right) \times (1+z)_2^{1.04} \times  (sSFR/sSFR_{MS})^{0.70} \times  M_{*~5} ^{0.01} \right]$ \\
\\
$ \dot{\rm M}_{acc} = \rm 73~\msun~yr^{-1}  \times   \left[2.3\times (1+z)_2^{3.60} \times  \left( M_{*~5}^{0.56}  -0.56 \times M_{*~5} ^{0.74}\right)\right] $ \\
\\ 
\\
$ \rm \tau_{dep} \equiv \rm {M_{\rm ISM} \over{SFR}} = \rm 1.01 ~ Gyr~\times \left[ \rm (1+z)_2^{-1.04} \times  (sSFR/sSFR_{MS})^{-0.70} \times  M_{*~5}^{-0.01}\right]$     \\
\\
$ \rm gas/stellar \equiv \rm {M_{\rm ISM} \over{M_{\rm *}}} = \rm 1.74~\times  \left[\rm (1+z)_2^{1.84} \times  (sSFR/sSFR_{MS})^{0.32} \times  M_{*~5}^{-0.70}\right]$     \\
\\
$ \rm f_{gas} \equiv \rm {M_{\rm ISM} \over{ M_{\rm *} + M_{\rm ISM} }} =   0.63~\times \left[1.58/ \left(1 ~+~0.58\times \rm (1+z)_2^{-1.84} \times  (sSFR/sSFR_{MS})^{-0.32} \times  M_{*~5}^{0.70} \right)\right]$ \\
\\
\hline
\hline
\end{tabular}
\caption{ISM, SFR and accretion normalized. The equations are written in a form such that the quantity in $\left[~\right]$ in each equation is equal to unity at z = 2 and $M_{*} = 5\times10^{10}$\msun. $\rm (1+z)_2$ is $(1+z)$ normalized to its value at z = 2 where (1+z)=3. M$_{*~5}$ is the stellar mass normalized to $5\times10^{10}$\msun.  The fourth relation is obtained by canceling out the M$_{\rm ISM}$ term in the second equation, not by division of the first equation by the second. See original equations in text for the uncertainties in the coefficients.}
\label{equations2}
\end{table}


\begin{figure}[ht]
\includegraphics[width=0.48\textwidth]{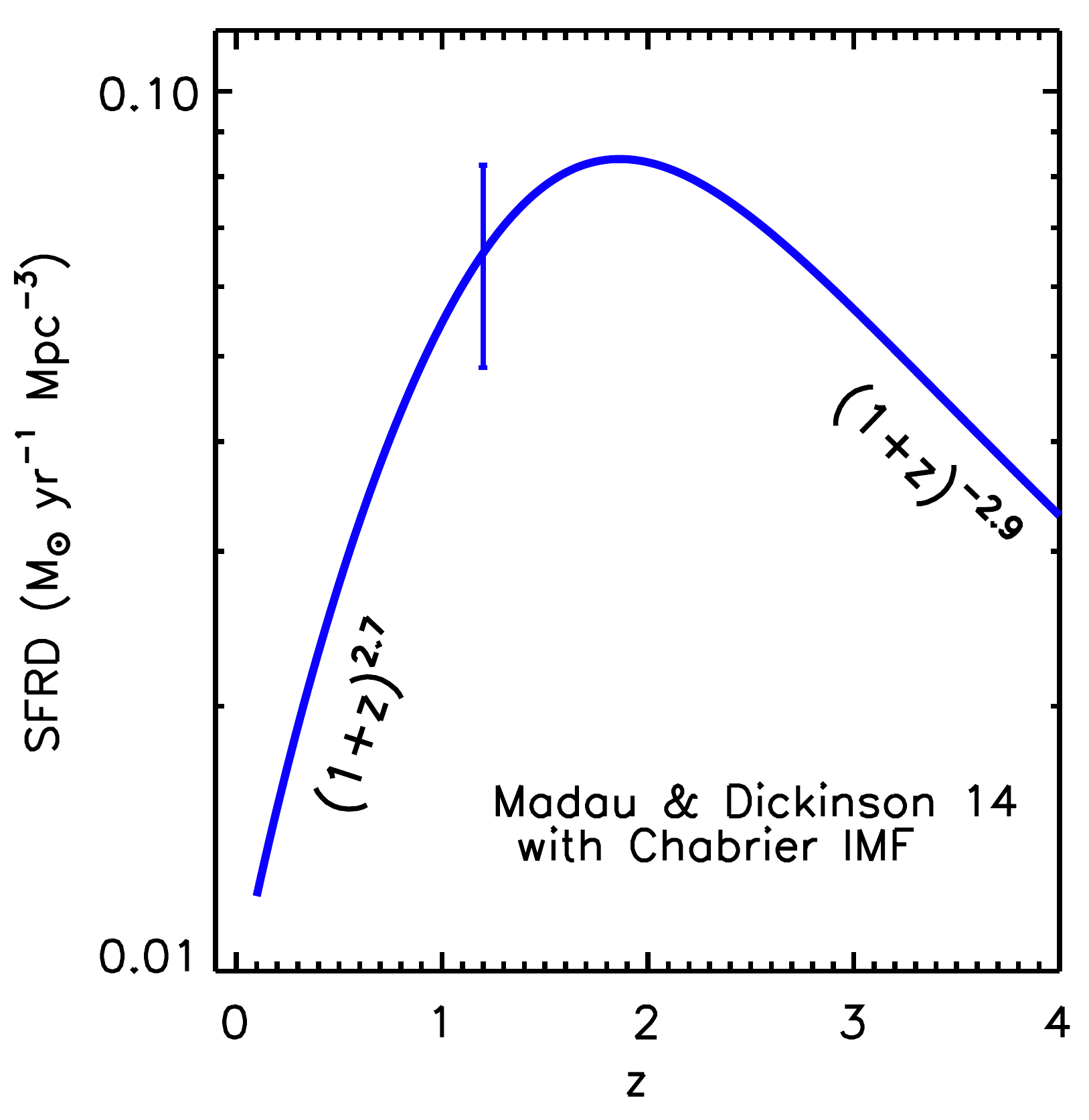}
\includegraphics[width=0.48\textwidth]{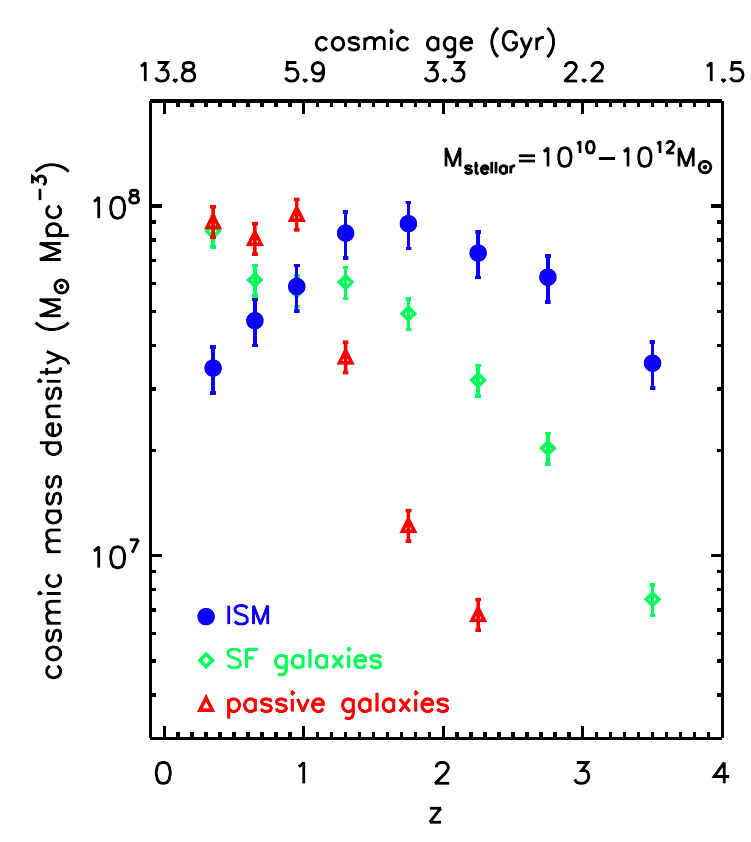}
\caption{The evolution of the cosmic star formation rate density (SFRD) from \cite{mad14} is shown for comparison with the overall evolution of ISM content derived from the 
dust emission (Right panel, \cite{sco17}). The SFRD evolution has been corrected to 
a Chabrier IMF from the original Salpeter IMF used in \cite{mad14}.
A typical uncertainty in the SFRD is shown as 0.1 dex \citep[see][]{mad14}. }
\label{mad_dickinson} 
\end{figure}

The gas mass fractions computed for galaxies with $M_{\rm *} = 10^{10}$ to $10^{12}$ \msun~ are shown in Figure \ref{lilly_madau_2}. The ISM is dominant over the stellar mass down to z $\simeq 1.5$.  At $z = 3$ to 4 the gas mass fractions 
get up to $\sim 80$\% when averaged over the galaxy population. Thus, the evolution of ISM contents which peak at $z \simeq 2$, is likely responsible for the peak in SF and AGN activity at that epoch \citep[and references therein]{mad14}. At $z = 4$ down to 2, the buildup in the ISM density is almost identical to that of the cosmic SFRD shown in Figure \ref{mad_dickinson}. 

\section{Cross-Correlations with the Kinematic Sunyaev-Zel'dovich Effect}

While the synergies with SPHEREx and large-scale Cosmic Microwave Background (CMB) surveys are numerous, cross-correlation studies with the kinematic Sunyaev-Zel'dovich effect are among the most promising. CMB photons interact with late-time matter through a number of scattering and gravitational processes that imprint additional small-scale anisotropy on the CMB.  
In particular, the thermal and kinematic Sunyaev-Zel'dovich effects (tSZ and kSZ), caused by inverse Compton scattering of CMB photons off $kT \sim 1 \,$keV electrons, allow us to directly image the gas near the outskirts of galaxies and can yield important information on the thermodynamics and formation history of massive halos.  These SZ probes are particularly interesting because the amount of ionized gas observed inside galaxies and clusters falls short of the cosmological abundance -- the so-called ``missing baryon problem'' \cite{2004ApJ...616..643F}. 
The majority of this gas is expected to be located in the outskirts of galaxies in a diffuse state,
making it difficult to observe directly, especially for low mass halos at high redshifts.  Locating these missing baryons in galaxy groups would have far-reaching implications for galaxy formation models by helping constrain the amplitude and effect of feedback processes \cite{2017JCAP...11..040B}.

Currently, the diffuse outskirt gas is poorly characterized, which is problematic for a variety of cosmological measurements.  For example, taking full advantage of upcoming large-scale structure experiments like Euclid, LSST, and WFIRST will require modeling the matter power spectrum to percent precision \cite{2015ApJ...806..186O, 2015MNRAS.454.2451E}.
Since baryons constitute more than $15\%$ of the total mass in the universe, the current uncertainty on the baryon profile in halos is therefore a major limiting systematic in such studies. 

The kSZ effect, which is sourced by the Doppler boosting of CMB photons scattering off electrons with a non-zero peculiar velocity, offers a way to study this gas.  In temperature units, the shift $\Delta T^{\rm kSZ}(\bm{\hat{n}})$ produced by the kSZ effect is sourced by the free electron momentum field  $n_e \bm{v}_e$, and is given by \cite{sun72, 1986ApJ...306L..51O}:
\be
\frac{\Delta T^{\rm kSZ}(\bm{\hat{n}})}{T_{\rm CMB}}  = -  \sigma_T \int \frac{d \chi}{1+z} e^{-\tau(\chi)} n_e(\chi\hat{\bm{n}},\chi) \ \frac{\bm{v}_e}{c} \cdot \bm{\hat{n}},
\label{eq:kSZdef}
\ee
where $\sigma_T$ is the Thomson scattering cross section, $\tau$ is the optical depth to Thomson scattering, $\chi(z)$ is the co-moving distance to redshift $z$, $n_e$ and $\bm{v}_e$ are the \textit{free} electron number density and peculiar velocity, and $\bm{\hat{n}}$ is the line-of-sight direction.  The kSZ signal is linear in the gas mass, 
making it a unique probe of the low density and low temperature regions expected in the galaxy outskirts.
The kSZ measurements can be compared with most other observables, either non-linear in the gas mass or dependent on the gas temperature, to search for and constrain the formation and evolution of massive halos.

We forecast the expected signal-to-noise ratio $S/N$ of kSZ measurements combined with photometric redshifts from SPHEREx following the formalism of \cite{2017JCAP...11..040B}, which includes realistic gas profiles and CMB noise properties.
The kSZ $S/N$ is mostly sensitive to the actual electron profile, which is uncertain for low mass galaxy groups, and to the noise in the CMB temperature maps, including from the atmosphere. 
Most kSZ estimators in the literature \cite{2009arXiv0903.2845H, 2011MNRAS.413..628S, 2014MNRAS.443.2311L, 1999ApJ...515L...1F, 2008PhRvD..77h3004B, 2015arXiv151006442S} require either spectroscopic redshifts or very accurate photometric redshifts ($\sigma(z) \lesssim 0.01$).  For the most conservative estimate, we only include the two samples with highest redshift accuracy from \cite{2014arXiv1412.4872D}. 
We assume a bias $b \approx 1.1$, redshift $\bar{z} \approx 0.3$, a total number of galaxies on half of the sky $N \approx 24.5$ million,
and a CMB map noise of 7 $\mu$K-arcmin, typical of an upcoming ``Stage-3'' wide-field experiment.
The resulting kSZ $S/N$ is $70$, much higher than current measurements ($S/N\sim 3$).
Future CMB experiments, such as a proposed ``CMB S4'' survey \cite{2016arXiv161002743A}, are expected to have smaller map noise and should allow an even higher $S/N$ detection, even though the amplitude of the noise from fluctuations in the atmosphere is still uncertain.

A technique that operates in the absence of detailed redshift information has been developed \cite{2004ApJ...606...46D}. 
With this method, only a statistical redshift distribution for the overall source sample is required, which would allow us to use the full SPHEREx catalog. 
The concept is to cross-correlate tracers with the square of an appropriately filtered CMB map. 
The squaring operation circumvents the cancellation of the kSZ signal due to the alternating signs of the line-of-sight velocities, thus allowing a detection in cross-correlation.
This estimator has been recently used to detect the kSZ signal from WISE-selected galaxies in combination with Planck CMB data \cite{2016arXiv160301608H}. 
Forecast work \cite{2016PhRvD..94l3526F} suggests that a very large signal to noise ($S/N \gtrsim 100$) can be achieved by combining SPHEREx with upcoming CMB experiments. 
This method requires a very good foreground control on the CMB temperature map, and may ultimately be limited by residual contamination. 

As we motivate in the preceding discussion, the combination of current and future CMB experiments with large-scale structure surveys like SPHEREx will yield very high significance detections of the kSZ effect. This, in turn, will allow precision measurement of the baryon profile in the outskirts of the SPHEREx galaxies.  The achieved $S/N$ are sufficient to allow, for the first time, detailed sub-sample studies.  For example, 
the baryon profile is expected to depend both on the mass and the redshift of the host halo, and could depend on other galaxy properties. The unique spectral coverage of SPHEREx over the full sky allows for the selection and comparison of different populations, shedding light on the effect of feedback and star formation on the gas. Moreover, when combined with tSZ and lensing measurements, a full thermodynamic picture of the host halo can be obtained, constraining the amount of energy injected by feedback, as well as the gas density, temperature and the fraction of non-thermal pressure support \cite{2017JCAP...11..040B}.  Related techniques can be used to turn the kSZ measurements in constraints on the velocity correlation function, which is a potentially powerful probe of scale-dependent modified gravity.

\section{Novel Scientific Opportunities Enabled by SPHEREx}

In this section, we discuss novel scientific opportunities that were presented at this community workshop.

\subsection{Other probe of primordial non-Gaussianity with unbiased tracers}

\begin{figure}
 \begin{center}
    \includegraphics[width=0.6\columnwidth]{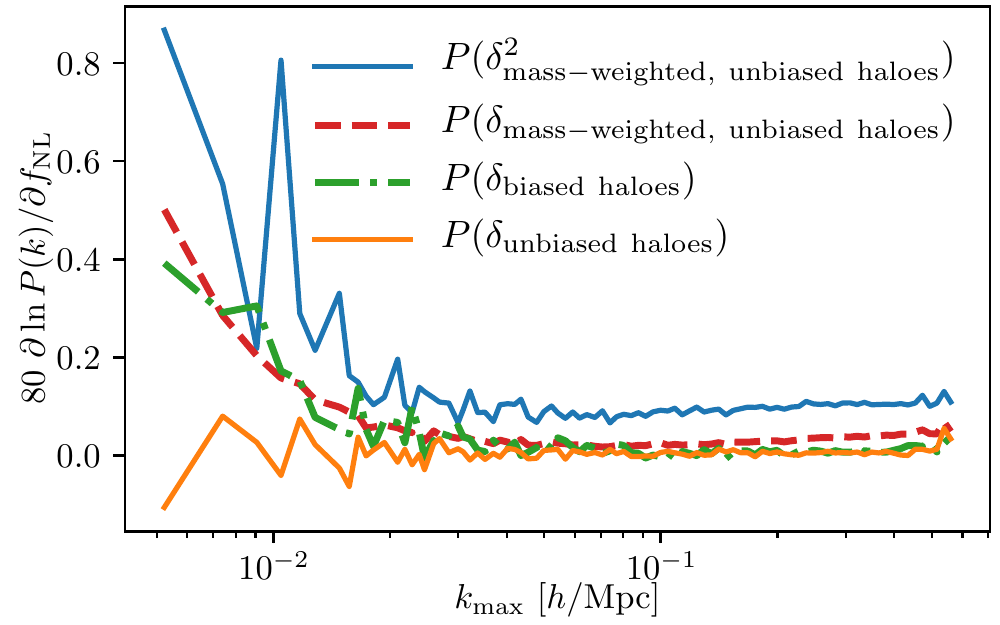}
  \end{center}
	\caption{Producing a sensitivity to $f_{\rm NL}$ from an unbiased tracer for which some mass information is available. The vertical axis shows the change in the log of the power spectrum when $f_{\rm NL}$ changes from 0 to 80 in a (1.2 $h^{-1}$Gpc)$^3$ volume; error bars on $f_{\rm NL}$ scale inversely with this quantity. This large-scale bias was estimated using cross-correlations with the matter density field.}
  \label{fig:fnl_from_unbiased}
\end{figure}

The standard approach to detect $f_{\rm NL}$ from galaxy surveys is to use highly biased tracers, relying on nature to do the biasing. This sensitivity to $f_{\rm NL}$ comes by increasing the weight in collapsed regions, tending to be in large-scale potential wells if $f_{\rm NL}>0.$ But instead of relying on nature, we can also bias the field ourselves, as explored by Neyrinck, et al. (2018, in prep.). This works best if we have some estimate of halo mass, but a proxy such as luminosity may work, as well. Fig.~\ref{fig:fnl_from_unbiased} shows the effect on various power spectra from changing $f_{\rm NL}$ from 0 to 80, measured from $1.2\ h^{-1}$Gpc, $1024^3$-particle simulations \citep{Pillepich:2010}. The power spectrum of `unbiased' halos of mass $>\sim 10^{13} M_\odot$ is essentially insensitive to $f_{\rm NL}$. The power spectrum becomes sensitive to $f_{\rm NL}$ when the mass threshold increases by a factor of 10, in the `biased' sample. But similar sensitivity comes through mass-weighting the halos of the $10^{13} M_\odot$ sample. This signal can even be enhanced by further increasing the weight of high densities, i.e. applying a Box-Cox \citep{Box:1964,Joachimi:2011} transform with index 2 (top curve). Many details remain to be worked out: whether to use an estimate of mass, or luminosity, to weight the galaxies or halos, and what transformations will give the best results in light of various noise sources. But in principle, this introduces a whole family of estimators of $f_{\rm NL}$, even from a single galaxy sample.

SPHEREx is ideally suited to make these measurements, not only from the biased tracers being used for the main $f_{\rm NL}$ constraints, but from other, less-biased tracers it can characterize. In particular, SPHEREx's unprecedented map of the extra-galactic background light  will allow an estimate of $f_{\rm NL}$ highly complementary to that from the standard probe.

\subsection{Probing the three phases of Hydrogen with SPHEREx at High Redshift} 

SPHEREx can probe three states of hydrogen in galaxies: (\textit{i}) shock-heated H2 via the vibrational line emission at $\lambda_{rest} \simeq 2.1\ \mu$m; (\textit{ii}) ionized HII associated with active star-forming (SF) regions and starburst galaxies in H$\alpha$ ($\lambda_{rest} \simeq 6563 \,$\AA) (out to $z \sim 3$); and (\textit{iii}) cosmic HII structures during the epoch of reionization (at $z \sim 6 {-} 10$) in Lyman-$\alpha$ emission. In each case below, we express the expected emission in terms of the line luminosity in solar units (\lsun).

\subsubsection{H$_2$ Emission}
The 2.12$\mu$m H$_2$ $v = 1-0$ emission lines are a major probe of shock-cooling molecules behind shocks at velocities of $5 - 40$~\kms. As this gas cools from 
$4000$ K to $1000$ K, this band is responsible for the dominant cooling. Very approximately, we can expect $\frac{5}{2} ~k \Delta T$ ergs of energy radiated per molecule behind the shock, where $k$ is the Boltzmann constant and $\Delta T = 3000$ K.  If on average every molecule going into star formation passes though such a shock once, then the H$_2$ cooling luminosity will be:
\begin{equation}
\rm L_{\rm H2}  \sim \rm 5\times10^5 ~ \left[SFR/(100 \, \msun {\rm yr}^{-1})\right] \lsun.
\end{equation}  
Using the 5$\sigma$ (all-sky) sensitivity for SPHEREx in the 2$\mu$m band, this implies the ability to detect a SFR $\sim 10 \, \msun$ yr$^{-1}$ per pixel at a distance of 20 Mpc. 

In addition to the crude physical estimate above, the H$_2$ luminosity can be derived empirically using existing observations of low-$z$ galaxies. In a sample of ULIRGs Davies et al.~\citep{Davies:2003} find the intensity of the H$_2$ S(1) line at 2.12$\mu$m to be generally comparable to that of the HI Br$\gamma$ line. Using Case B recombination line intensities for H$\alpha$ (that is, assuming that photons above 13.6 eV are reabsorbed; see Eqn.~\ref{eqn:ha} below) and assuming that Br$\gamma$ is 104 times weaker than H$\alpha$ (\textit{e.g.}, \citep{Osterbrock:1989}), we find:
\begin{equation}
\rm L_{\rm H2}  \sim \rm 5\times10^7 ~ \left[SFR/(100 \, \msun {\rm yr}^{-1})\right] \lsun. 
\end{equation} 
This line should be detectable in large numbers of local star-forming galaxies, and even some high-redshift star-forming galaxies.
 
 \subsubsection{H$\alpha$ Emission to probe SFRs}
 To assess the detectability of H$\alpha$ with SPHEREx we can use the well-developed relation between the 
 SFR and the Lyman continuum emission rate associated with the young stellar population, assuming a standard Chabrier IMF: 
 \begin{equation}
 \rm Q_{\rm Ly c}  = 1.4\times 10^{55} sec^{-1}  \left[SFR/(100 \, \msun {\rm yr}^{-1})\right], \label{eqn:Lyc} 
 \end{equation} 
and assuming Case B recombination, 
  \begin{equation}
 \rm L_{\rm H\alpha}  = 4.9\times 10^{9} \lsun  \left[SFR/(100 \, \msun {\rm yr}^{-1})\right].  \label{eqn:ha}
 \end{equation} 
 For the SPHEREx 5$\sigma$ (all-sky) sensitivities for H$\alpha$ as a function of redshift, 
 the implication is that we should detect any SFR regions with rates
  $>20 \, \msun$ yr$^{-1}$ at $z = 0.14$ to $3000 \, \msun$ yr$^{-1}$ at $z = 6.6$. The latter are 
  probably about the maximum to be expected at the end of reionization. At intermediate redshifts 
  one can expect to detect SFRs like those in ULIRGs, provided that the extinctions are not prohibitive.  

 \subsubsection{HI Lyman-$\alpha$ during Epoch of Reionization (EOR)}
 
 To estimate the expected fluxes in Lyman-$\alpha$ from galaxies and large scale structure 
 ionized bubbles during EOR ($z = 6 {-} 10$), we take the most optimistic viewpoint, namely 
 that the HII emission can be modeled as Case A recombination (photons with energy above 13.6 eV are not re-absorbed) with no dust, which is probably 
 valid outside individual galaxies. 
 Then assuming Eqn.~\ref{eqn:Lyc} again, we have:
  \begin{equation}
 \rm L_{\rm Ly\alpha}  = 3.7\times 10^{10} \lsun  \left[SFR/(100 \, \msun {\rm yr}^{-1})\right],   
 \end{equation} 
which predicts a factor of $\sim10$ greater than the relation for H$\alpha$. 
 
 At $z = 5.16$, the lowest redshift of Ly$\alpha$ covered by the SPHEREx bands, the co-moving volume per resolution element is $\sim$ 400 cMpc$^3$. At this redshift, the EOR models of Trac et al. \citep{Trac:2006} have an average co-moving SFR $\sim 0.1 \, \msun$ yr$^{-1}$, which yields a luminosity in Ly$\alpha$ similar to the SPHEREx 5$\sigma$ (deep) sensitivity. At EOR the structures are likely to be highly biased, and therefore a large fluctuation in star formation activity from place to place can be expected. This would imply that the there are likely to be areas where 
 the individual resolution elements are detectable without resort to stacking or two-point statistical techniques such as correlation functions.
 
\subsection{Forecasts for the number of detectable QSOs at $z>6$ in SPHEREx}

Quasi-Stellar Objects (QSOs) are among the most luminous objects in the Universe. They are triggered by the fall of gas onto a super-massive black hole (SMBH), and it is believed that every galaxy has one at its center. Thanks to their high luminosity, QSOs can be used to trace the large-scale structure of the Universe up to very high redshift (high-$z$) using their clustering \citep{Ata2018} or Lyman $\alpha$ forest observations \citep[e.g.,][]{Font-Ribera2013}. In addition, the number of QSOs at $z>6$ can be used to set constraints on the epoch of reionization \citep[e.g.,][]{Fan06}.

To predict the number of QSOs detectable by SPHEREx at high-$z$ ($z>6$) we analyze the cosmological hydrodynamical simulation {\sc BlueTides} \citep{Feng16}. {\sc BlueTides} evolved $2\times7040^3$ particles in a comoving volume of $400^3\,h^{-3}$Mpc$^{3}$ from $z=99$ to $z=7.6$. We find this simulation especially suited to do forecasts for SPHEREx, as the main source of uncertainty in high-$z$ predictions is cosmic variance and {\sc BlueTides} is the largest high-$z$ simulation up-to-date. 

\begin{figure}
\begin{center}
\includegraphics[width=0.475\textwidth]{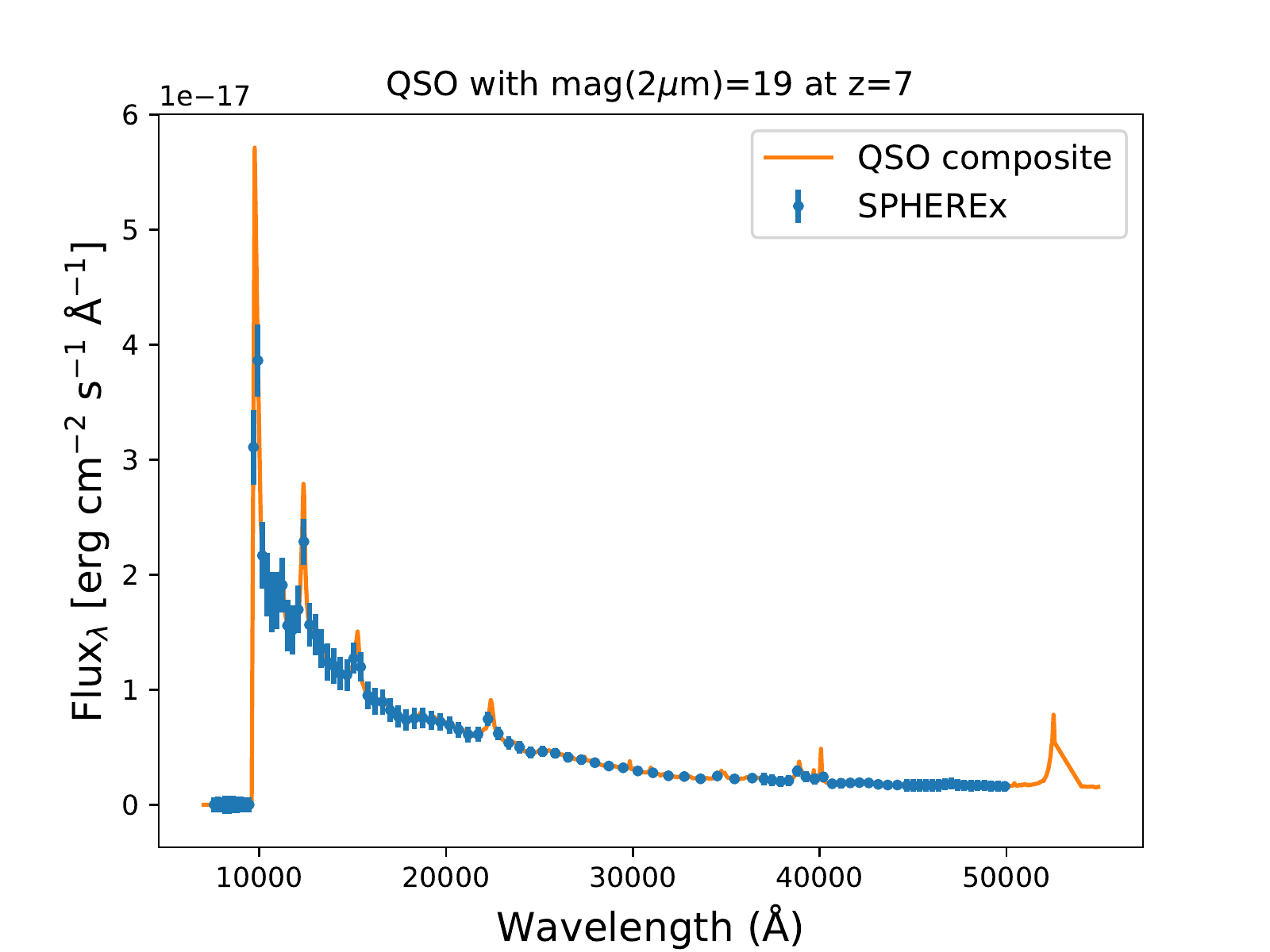}
\includegraphics[width=0.475\textwidth]{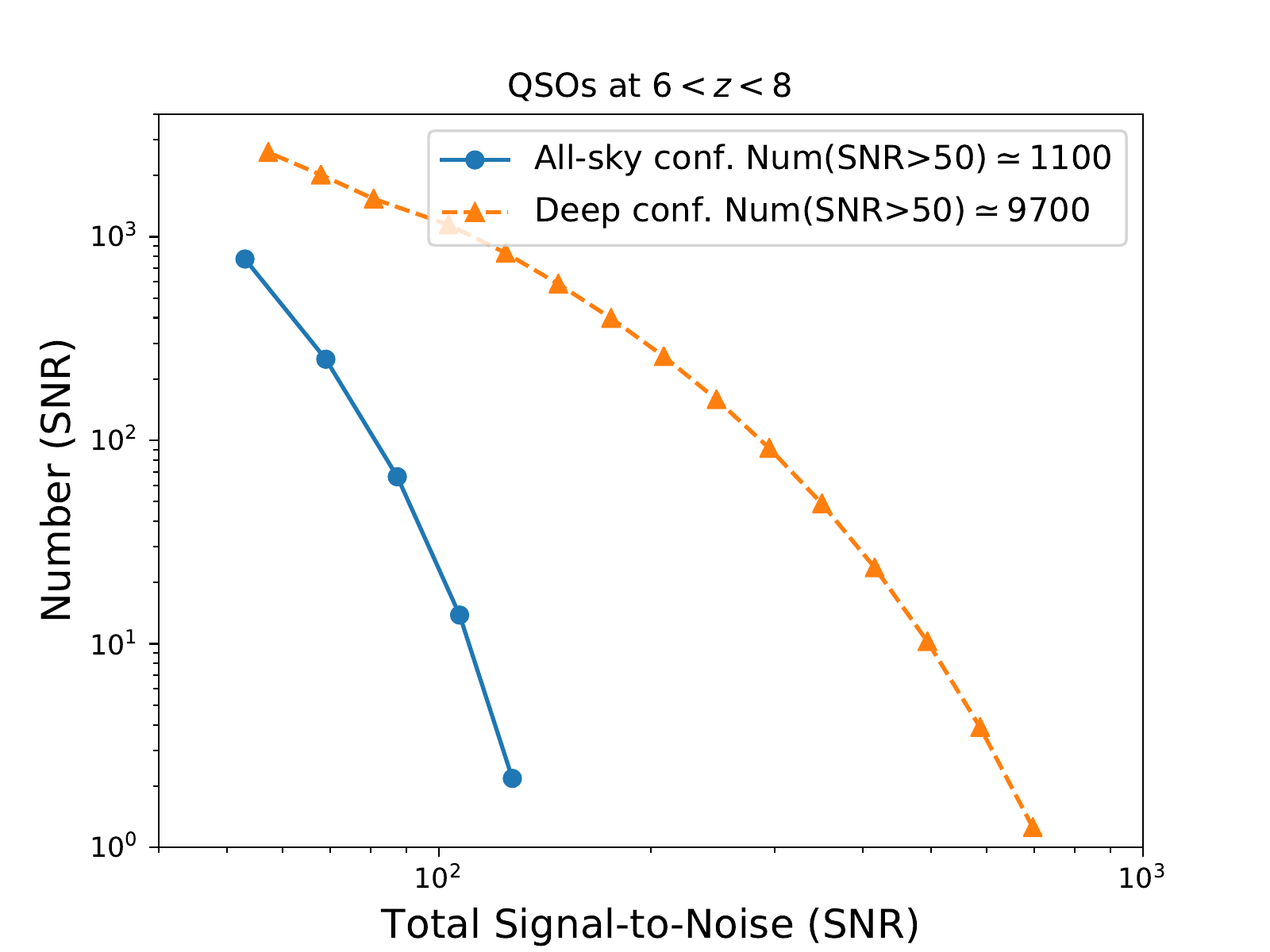}
\end{center}
\caption{
\label{fig:qso_num}
{\bf Left panel}: spectral energy distribution of a QSO with magnitude mag$(2\mu {\rm m})=19$ at redshift $z=7$ (orange line) and simulated SPHEREx observation (blue dots). The spectral resolution of SPHEREx allows to resolve multiple QSO emission lines, which enables to unambiguously identify QSOs and to compute their redshifts with great precision.
{\bf Right panel}: number of QSOs at $6<z<8$ detectable by SPHEREx and SPHEREx-deep as a function of detection signal-to-noise. SPHEREx will detect the $1\,100$ brightest high-$z$ QSOs in the Universe and $10^4$ additional QSOs in the polar regions.
}
\end{figure}

In the left panel of Fig.~\ref{fig:qso_num} we display the spectral energy distribution (SED) of a QSO with magnitude mag$(2\mu {\rm m})=19$ at redshift $z=7$ (orange line) and a simulated SPHEREx observation of this object (blue dots). The SED is computed by applying extinction shortward Lyman $\alpha$ due to neutral hydrogen at high-$z$ \citep[using Gun-Peterson opacity measurements from][]{Barnett2017} to a QSO composite created from the spectra of $2\,200$ SDSS QSOs at $z<5$ \citep{VandenBerk2001}. The spectral resolution of SPHEREx is more than enough to resolve several QSO emission lines, which can be used to compute QSOs redshifts with very good precision and to identify them with great reliability using automatic algorithms \citep{Chaves-Montero17}.

In the right panel of Fig.~\ref{fig:qso_num} we show the number of QSOs at $6<z<8$ that can be detected by SPHEREx (blue line) and SPHEREx-deep (orange line) as a function of detection signal-to-noise (SNR). We employ a Schechter function to model the number of QSOs as a function of observed magnitude, and we determine this function by fitting the number of QSOs in {\sc BlueTides} at $z=7.6$. We note that a Schechter function provides a lower limit for the number of bright sources, as {\sc BlueTides}' data is also well fit by a power law. To make predictions at different redshifts we account for the evolution in the number of QSOs using Fig.~5 of \citet{Waters2016}. In the $y$-axis we show the total detection SNR, which is computed by combining that of every frequency unit with SNR$>0.5$ for simulated SPHEREx observations. We expect SPHEREx to detect the $\simeq 1\,100$ most luminous QSOs in the Universe with a total SNR$>50$, which will set the strongest constraints on the bright end of the QSO luminosity function up to redshift $z=8$. Furthermore, in the polar regions SPHEREx will detect around $\simeq 10^4$ QSOs, which will allow to systematically study the redshift evolution of the fraction of neutral Hydrogen during the epoch of reionization.

\section{Conclusion}

The two SPHEREx Community Workshops identified and discussed many important scientific investigations enabled by the unique 0.75-to-5 $\mu$m all sky spectral data base which would be the legacy of SPHEREx. The combination of this Legacy Science with SPHEREx’ expected major contributions to the three science themes which drive the mission design would allow SPHEREx to follow in the illustrious scientific footsteps of earlier all-sky surveys such as IRAS, COBE, Planck and WISE and make SPHEREx a worthy participant in the ``Decade of the Surveys''.  The 2018 Workshop, reviewed and summarized in this White Paper, concentrated on the many synergies between SPHEREx and other missions and data bases, including predecessor missions such as WISE and TESS, contemporaneous missions such as JWST and Euclid, and even successor missions, including WFIRST and PlATO.  The synergies extend beyond SPHEREx VIS/IR spectral band to include the X-ray mission eROSITA and the ALMA sub-millimeter array, as additional examples.  

The coupling of SPHEREx to NASA’s next major space observatory, JWST, is particularly important because the JWST wavelength range, measurement capabilities, and schedule encompass those of SPHEREx.  However, SPHEREx all-sky survey capability adds a dimension which JWST cannot approach.  On the current schedule, in which JWST launches in mid-2020 and SPHEREx in 2023, the first SPHEREx data releases, publicly available in mid-2023, can be mined for puzzling and unusual targets ${-}$ candidate high redshift quasars and sources with unusual ice spectra, to cite two possible examples ${-}$ which can be further explored with JWST’s superior sensitivity, and spectral and spatial resolution.  Delaying SPHEREx beyond 2023 would undermine this valuable synergy.  

The scientific examples discussed throughout this White Paper illuminate the power of this all-sky survey, which will provide spectra of hundreds of millions of objects.  The SPHEREx data would be a new tool for astronomers, but the previously described examples are necessarily limited to at most an extension of our current knowledge.  As a final thought, we point out that SPHEREx, like all sky surveys which open new parameter space, has the potential to yield new discoveries which change our view of the astronomical Universe.

\acknowledgments
O.D and M.W. would like to warmly thank Gary Melnick, Matt Ashby and Volker Tolls for leading the logistical effort that made this workshop so successful. Part of the research described in this paper was carried out at the Jet Propulsion Laboratory, California Institute of Technology, under a contract with the National Aeronautics and Space Administration. The information presented about SPHEREx is pre-decisional and is provided for planning and discussion purposes only.

\bibliographystyle{JHEP}

\end{document}